\documentclass[12pt]{article}

\usepackage{amsmath, amsfonts, amssymb}  
\usepackage{graphicx}  
\usepackage{cite}  
\usepackage{url}  
\usepackage{geometry}  
\usepackage{titlesec}  
\usepackage{lipsum}  
\usepackage{setspace}  
\usepackage{abstract}  
\usepackage{enumitem}  
\usepackage{booktabs}  
\usepackage{caption, subcaption}  
\usepackage[all]{nowidow}  
\usepackage[utf8]{inputenc}  
\usepackage{rotating}  
\usepackage{appendix}  
\usepackage{bm}  
\usepackage{tikz}  
\usetikzlibrary{er, positioning, bayesnet}  
\usepackage{multicol}  
\usepackage{algpseudocode, algorithm, algorithmicx}  
\usepackage{listings}  
\usepackage{hyperref}  
\usepackage{longtable}  
\usepackage{array, colortbl, xcolor}  
\usepackage{multirow}  
\usepackage{amsthm}  
\usepackage{mathrsfs}  
\usepackage{manyfoot}  
\usepackage{float}  
\usepackage{placeins}  
\usepackage{adjustbox}  
\usepackage{eqparbox}  
\usepackage{soul}  
\usepackage{icomma}  
\usepackage{marvosym}  
\usepackage{tabularx}  

\title{\textbf{Decoding Financial Health in Kenya’s Medical Insurance Sector: A Data-Driven Cluster Analysis}}

\author{Evans Kiptoo Korir\thanks{University of Szeged, Hungary} \and Zsolt Vizi\thanks{National Laboratory for Health Security}}
\date{\today}

\titleformat{\section}
{\normalfont\Large\bfseries}{\thesection}{1em}{}
\titlespacing*{\section}{0pt}{\baselineskip}{\baselineskip}

\titleformat{\subsection}
{\normalfont\large\bfseries}{\thesubsection}{1em}{}
\titlespacing*{\subsection}{0pt}{\baselineskip}{\baselineskip}

\providecommand{\keywords}[1]{\textbf{\textit{Keywords---}} #1}

\begin{document}

\maketitle

\begin{abstract} This study explores the financial performance and reporting trends of insurance companies in the medical sector using advanced clustering techniques to identify distinct patterns. By analyzing financial ratios and time series data, four unique clusters were identified, each representing different combinations of financial stability and reporting consistency. To capture temporal variations and uncover key behavioral insights, Dynamic Time Warping (DTW), K-Means, and Hierarchical Clustering methods were employed.
The results highlight that insurers with stable financial performance and consistent reporting practices emerge as reliable options for policyholders. Conversely, clusters containing underperforming companies and those with reporting gaps reveal significant operational challenges and data consistency issues. These findings underscore the critical role of transparency and timely financial reporting in maintaining the sector’s resilience.
This study contributes to the literature by integrating time series analysis into financial clustering, offering practical recommendations for enhancing data governance and financial stability in the insurance industry. Future research could further explore non-financial indicators and alternative clustering approaches to gain deeper insights into performance dynamics.
\end{abstract}


\section{Introduction}

The insurance sector is a cornerstone of the Kenyan economy, driving resilience and economic stability while fostering growth. Over the years, the industry has made significant progress, with the Insurance Regulatory Authority (IRA) reported a gross written premium of KES 361.36 billion by the end of Q4 2023, a 16.7\% increase from KES 309.77 billion in Q4 2022. However, challenges persist. The general insurance business recorded underwriting losses that increased from KES 3.72 billion in Q4 2022 to KES 4.96 billion in Q4 2023, while claims paid increased by 12.3\% to KES 81.16 billion. These trends underscore the industry’s growth and challenges, particularly in managing claims effectively and maintaining profitability \cite{ira}.

At the same time, policyholder needs for reliable medical insurance coverage are intensifying, driven by Kenya's commitment to universal health coverage (UHC) by 2030 \cite{WHO}. The government has implemented numerous health reforms, such as expanding the National Hospital Insurance Fund (NHIF) to cover outpatient services and subsidizing premiums for vulnerable households \cite{Kazungu, Okech}. Despite these efforts, Kenya’s health financing system remains fraught with challenges. High out-of-pocket payments, which account for 27\% of total health expenditure, expose citizens to financial risks, with over 453,470 individuals falling into poverty annually due to healthcare costs \cite{Barasa}. This context underscores the critical need for robust, affordable medical insurance solutions to protect policyholders against economic shocks.

In response to these challenges, Kenya recently transitioned from NHIF to Taifa Care, a rebranded Social Health Insurance Fund (SHIF). Established under the Social Health Insurance Act of 2023, Taifa Care aims to provide equitable healthcare coverage by mandating income-based contributions. However, its rollout has faced resistance due to concerns about affordability, rushed implementation, and operational challenges. As Taifa Care seeks to offer universal coverage, it also highlights the role of private insurers in complementing government programs, filling coverage gaps, and providing tailored solutions to meet the needs of diverse policyholders.

In its various forms, insurance plays a pivotal role in mitigating financial risks. Life insurance safeguards against income loss due to unforeseen events, property insurance protects assets, and medical insurance helps cover healthcare costs. However, insurance companies face significant risks, such as credit, underwriting, solvency, liquidity, and reinsurance risks, that directly impact their performance. These risks are particularly pronounced in developing economies like Kenya, where regulatory frameworks may be less robust, limiting insurers’ capacity to manage risks effectively \cite{Omasete, Wilson, Baluch}. Poor risk management can exacerbate claims liabilities, hinder profitability, and erode policyholder trust.

Traditional methods for assessing financial health in the insurance industry often rely on individual financial ratios, which, while useful, fail to capture broader trends and interdependencies among key financial indicators. These static approaches may overlook crucial temporal dynamics that influence an insurer’s long-term viability. Clustering techniques offer a data-driven approach to grouping insurance companies based on their financial performance, revealing hidden patterns and emerging risks that might otherwise go unnoticed. By incorporating time-series analysis, this study enhances the ability to detect financial vulnerabilities, identify high-performing firms, and provide actionable insights for regulators, insurers, and policyholders.

This study applies clustering techniques, particularly Dynamic Time Warping (DTW) and Long Short-Term Memory (LSTM) networks, to analyze insurance firms based on financial ratios derived from publicly available data. The use of temporal clustering of quarterly time-series data allows for a more comprehensive evaluation of financial health, enabling comparisons over time and distinguishing firms based on their financial stability and reporting practices.
The findings of this study have implications for multiple stakeholders. Policymakers can leverage the insights to improve sectoral efficiency through regulatory reforms, insurers can identify strategic opportunities for improvement, and policyholders can make informed decisions when selecting providers. The study also contributes to the broader discourse on the intersection of financial performance, risk management, and sustainability in the insurance sector.

This paper is structured as follows: Section 2 provides an overview of the relevant literature. Section 3 details the methodology, including the research design, data sources, financial ratio selection, application of temporal feature fusion, clustering analysis, and cluster evaluation metrics. Section 4 reports the findings, followed by Section 5, which discusses the key results, addresses the study's limitations, and proposes avenues for future research. Finally, Section 6 concludes the paper. All analyses were performed in Python, using a variety of statistical and machine-learning libraries.

\section{Literature Review}
Health insurance plays a critical role in healthcare systems, particularly in developing countries like Kenya, where financial protection against rising medical costs is essential. Despite its significance, health insurance penetration in Kenya remains low, with the Insurance Regulatory Authority (IRA) reporting a penetration rate of only 2.25\% in 2021. Insurance density stood at 4,777 KES per capita, indicating limited participation in private health insurance programs. The Kenyan insurance sector, predominantly led by private providers, struggles with extending coverage to low-income and rural populations. These challenges are exacerbated by a high reliance on out-of-pocket payments, further straining financial security at the household level. While much of the existing research examines the financial health and operational efficiency of insurance providers, there remains a lack of comprehensive studies addressing the financial performance of health insurers in Kenya and their contribution to overall market stability.  

Prior studies on insurance markets have predominantly focused on traditional financial performance measures such as return on assets (ROA), return on equity (ROE), solvency ratios, and claims ratios. These metrics provide valuable insights into an insurer’s profitability, stability, and risk exposure. For instance, studies such as  \cite{Barasa, Kazungu, Korir2020, KorirArun, Korirthesis} analyze profitability trends in Kenya’s insurance sector using these indicators. However, these studies primarily assess financial ratios in isolation, rather than exploring their interrelationships or how they collectively shape financial sustainability. The fragmented nature of such analyses limits the ability to capture the dynamic behavior of insurers over time. In contrast, clustering techniques offer a more holistic approach by identifying financial patterns across multiple insurers, enabling segmentation based on shared financial characteristics. Despite the growing use of clustering in financial analysis, its application in assessing insurers’ financial health, particularly within Kenya’s health insurance sector, remains largely unexplored.  

Several studies outside the Kenyan context have applied clustering techniques to analyze financial performance. \cite{Wang} employed fuzzy clustering to group companies based on financial ratios, highlighting common financial characteristics among firms with similar performance. Similarly, \cite{Dzuba} used cluster analysis to classify firms by financial strategies, revealing distinct differences between high-tech and basic industries. \cite{Herman} extended this approach to food retail companies in Hungary and Romania, demonstrating the influence of methodological choices, such as K-Means versus K-Medoids on classification outcomes. \cite{Qian} introduced a financial K-Means algorithm for classifying companies in Zhejiang province, showing that clustering can be instrumental in assessing loan eligibility. While these studies underscore the utility of clustering in financial performance evaluation, they are largely concentrated in developed economies and industries outside the insurance sector. There remains an evident gap in adapting such methodologies to health insurance markets in sub-Saharan Africa, where financial performance is influenced by unique market dynamics, regulatory environments, and economic constraints.  

A notable limitation of prior clustering applications in finance is their reliance on static, one-time financial data rather than time-series financial ratios. The absence of a longitudinal approach makes it difficult to assess how insurers’ financial health evolves over time. Dynamic time-warping and advanced clustering techniques could provide deeper insights into temporal financial patterns, yet studies employing these methodologies in the insurance sector are scarce. Moreover, while existing research emphasizes profitability and solvency, limited attention has been given to how these financial ratios collectively influence the overall health of insurance markets in developing economies. The few studies conducted in Kenya, such as  \cite{Korir2020, KorirArun, Korirthesis, Mutua}, focus on select aspects of financial performance, such as claims incurred ratios and liquidity risk, but do not integrate a comprehensive analysis linking multiple financial indicators to market sustainability.  

The theoretical foundation for financial ratio analysis in insurance markets has been predominantly derived from financial distress prediction models and market efficiency theories. Traditional financial distress models, such as Altman’s Z-score \cite{Altman} and Merton’s Model \cite{Merton}, provide a basis for evaluating insurer solvency and predicting financial instability. Market efficiency theories suggest that financial ratios reflect underlying firm performance and risk exposure, influencing investor confidence and market stability. However, these models have rarely been adapted to explore financial clustering in insurance markets. By integrating clustering methodologies with financial distress prediction models, this study seeks to provide a novel framework for evaluating insurers’ financial health. The application of advanced clustering techniques can enhance understanding of financial stability trends, identifying insurers at risk of distress and offering insights into market competitiveness.  

Despite contributions from past research, there remain significant gaps in the literature. First, existing studies on financial performance in the Kenyan insurance market have not examined how financial ratios interrelate to influence overall market health. Most research has analyzed individual aspects, such as profitability or solvency, without considering how these dimensions interact within health insurance firms. Second, there is a lack of studies employing time-series clustering techniques to analyze financial performance evolution among insurers. Current approaches rely on static ratio assessments, which fail to capture long-term financial trajectories. Third, while international studies have demonstrated the potential of clustering for financial classification, its applicability to Kenya’s health insurance sector remains unexplored. The limited penetration of health insurance, coupled with regulatory and economic challenges unique to Kenya, necessitates a localized approach to financial performance analysis. Finally, critical financial indicators such as the Claims Payout Ratio, which reflects an insurer’s efficiency in handling claims, have not been adequately examined in the Kenyan context. Given the pivotal role of claims management in the sustainability of health insurers, understanding its interplay with other financial ratios is essential for evaluating market stability.  
This study addresses these gaps by employing a clustering approach to classify health insurers based on financial performance patterns, offering a more nuanced perspective on market health than traditional financial ratio analysis alone. By leveraging time-series data and advanced clustering methodologies, this research extends prior applications of financial clustering and contributes to the broader discourse on financial performance assessment in emerging insurance markets.

\section{Methodology}

\subsection{Research Design}

\begin{figure}[H]
    \centering
\includegraphics[width=\textwidth]{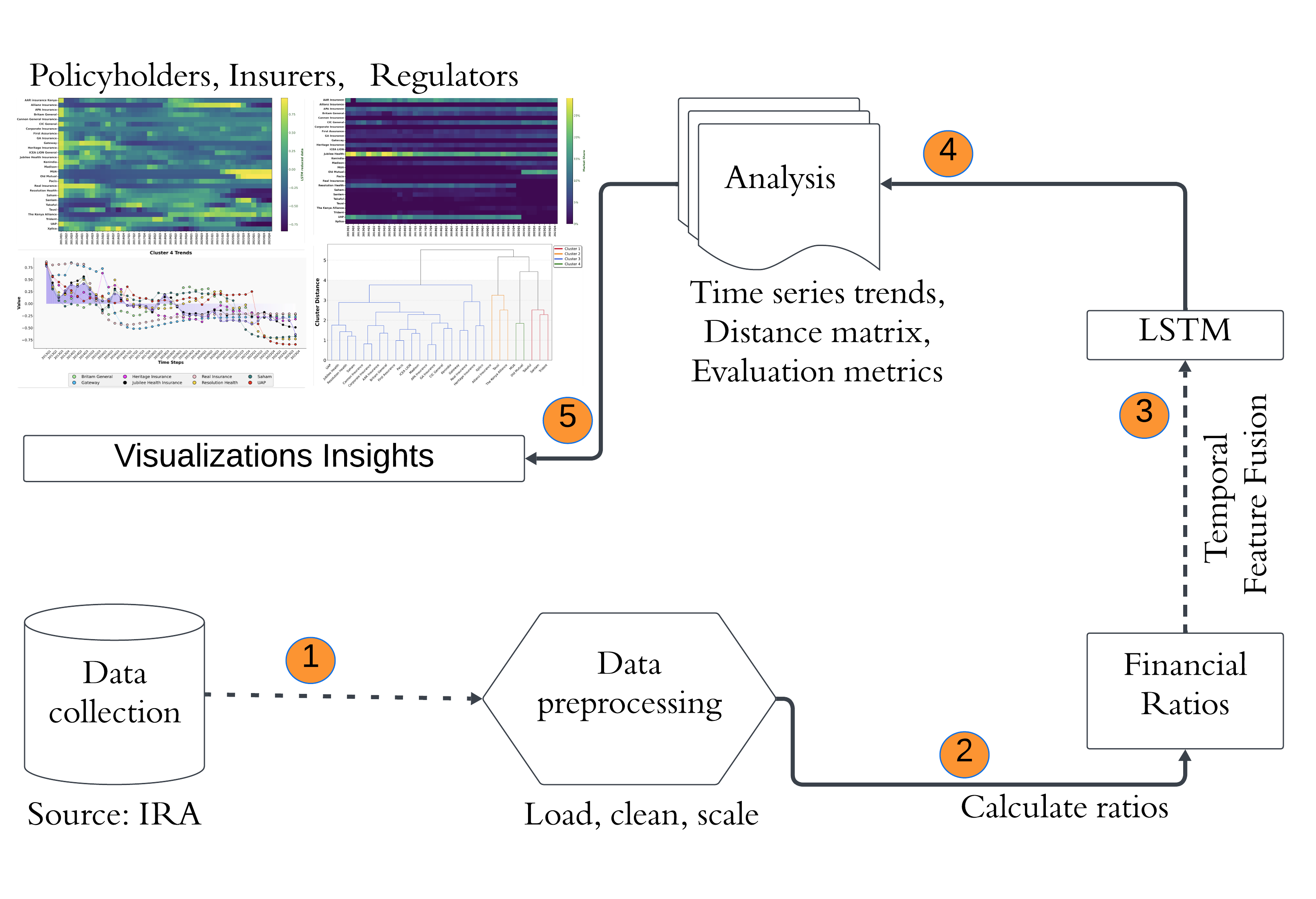}
    \caption{\footnotesize This flowchart illustrates the step-by-step process of analyzing medical insurance data in Kenya. It begins with data collection from the Insurance Regulatory Authority (IRA). It progresses through stages of data preprocessing, temporal feature fusion (LSTM), detailed analysis (including time series trends and DTW distance matrix), and visualization. The final output provides actionable insights and recommendations for policyholders and industry stakeholders, to improve decision-making in the insurance sector.    
    } 
    \label{fig: flowchart}
\end{figure}

This study employs a comprehensive research framework to analyze the performance of medical insurance providers in Kenya. The framework integrates various research procedures and analytical techniques to process and evaluate the data. The overall design is shown in Fig. \ref{fig: flowchart}, which illustrates the step-by-step process of data collection, preprocessing, temporal feature fusion, analysis, and visualization.

The research design integrates statistical and machine learning methods to extract meaningful insights from the data. Techniques such as Long Short-Term Memory (LSTM) for temporal feature fusion, Dynamic Time Warping (DTW) for analyzing temporal trends, and K-Means clustering for identifying patterns were employed to examine the financial performance of insurance providers.
These methods were implemented using Python's object-oriented programming framework, taking advantage of \texttt{Scikit Learn} and \texttt{clustering algorithms}. In addition, several open-source libraries were used to visualize financial ratios, trends, and other relevant metrics. The complete code for this study is publicly accessible on GitHub \cite{code}, promoting further development of the analysis framework.

\subsection{Data}

This study utilizes comprehensive financial data obtained from the Kenya Insurance Regulatory Authority (IRA), which publishes quarterly reports on the performance of insurance and reinsurance companies. The key data metrics analyzed in this study are summarized in Table~\ref{tab:data_metrics}.

\begin{table}[H]
\centering
\captionsetup{justification=centering} 

\arrayrulewidth=0.5mm 
\renewcommand{\arraystretch}{1.8} 
\small 
\setlength{\tabcolsep}{15pt} 

\begin{tabular}{l l}
\toprule
Data Metric & Description \\ \midrule
Gross Premium Income & Total revenue generated from insurance premiums from policyholders. \\
Claims Paid & Total amount disbursed to settle approved policyholder claims. \\
Claims Incurred & Total expected cost of claims during the period. \\
Underwriting Profit & Profit derived from insurance operations less claims and expenses. \\ 
\bottomrule
\end{tabular}

\caption{\footnotesize Key Financial Data Metrics Used in the Study. These metrics provide insights into the financial performance and operational efficiency of insurance companies.}
\label{tab:data_metrics}
\end{table}

These data, covering multiple fiscal years, were sourced directly from the IRA website \cite{ira}, ensuring both accuracy and adherence to regulatory standards.
No additional data collection methods, such as surveys or interviews, were utilized. To maintain data integrity, any missing values for companies were systematically recorded as zero. 
Comprehensive data preprocessing was conducted to ensure consistency and quality, making the dataset well-suited for robust statistical analysis.

\subsection{Financial Ratios}
This study employs key financial ratios to assess the performance and financial health of health insurance providers in Kenya. The selection of these ratios is based on established financial and actuarial research, aligning with prior studies on insurance firm performance, profitability, and risk exposure as reported in \cite{ira}. The chosen ratios provide a comprehensive assessment of operational efficiency, market position, and underwriting performance, which are critical in evaluating the sustainability and competitiveness of insurance providers.
A critical measure in this analysis is the Expense Ratio, which reflects the proportion of an insurer's operational costs relative to its income. It is calculated as the ratio of expenses to gross premium income. The expenses are estimated using the formula:
\[
\text{Expenses} = \text{Gross Premium Income} - (\text{Claims Incurred} + \text{Underwriting Profit})
\]
In addition to the Expense Ratio, the study evaluates other financial ratios, including Market Share, Claims Paid Ratio, Claims Incurred Ratio, Claims Payout Ratio, Underwriting Profit Ratio, and the Combined Ratio, as detailed in Table \ref{tab: Ratios}.
Extreme values were retained in the analysis to ensure meaningful variations, particularly those indicative of financial distress, are not suppressed. 
The financial ratios were normalized using the \texttt{StandardScaler}, standardizing each feature to have a mean of zero and a standard deviation of one. A \textit{within-company scaling} approach was used, where scaling parameters were computed separately for each company to preserve unique statistical characteristics.
\begin{table}[H]
\centering
\captionsetup{justification=centering}

\arrayrulewidth=0.5mm 
\renewcommand{\arraystretch}{1.8} 
\small 
\setlength{\tabcolsep}{12pt} 

\begin{tabular}{l l l}
\toprule
Ratio & Formula & Global Range \\ \midrule
Market Share & $\frac{\text{Premiums}}{\text{Total Market Premiums}} \times 100$ & 5\% - 30\%  \\ 
Claims Paid Ratio & $\frac{\text{Claims Paid}}{\text{Gross Premium Income}} \times 100$ & 50\% - 80\%  \\ 
Loss Ratio & $\frac{\text{Claims Incurred}}{\text{Gross Premium Income}} \times 100$ & 60\% - 75\%  \\ 
Underwriting Profit Ratio & $\frac{\text{Underwriting Profit}}{\text{Gross Premium Income}} \times 100$ & 5\% - 15\%  \\ 
Expense Ratio & $\frac{\text{Expenses}}{\text{Gross Premium Income}} \times 100$ & 10\% - 25\%  \\ 
Combined Ratio & $\text{Loss Ratio} + \text{Expense Ratio}$ & 90\% - 105\%  \\ 
Claims Payout Ratio & $\frac{\text{Claims Paid}}{\text{Claims Incurred}} \times 100$ & 60\% - 90\%  \\ 
\bottomrule
\end{tabular}
\caption{\footnotesize Key Financial Ratios used to evaluate insurance providers. The global range represents typical industry standards for each ratio, indicating the desired financial performance for insurance companies.}
\label{tab: Ratios}
\end{table}

\subsection{Long Short-Term Memory (LSTM)}
\label{LSTM}
Traditional financial analysis often relies on handcrafted ratios to capture key aspects of a firm's performance. However, financial data is inherently sequential, with past values influencing future performance. The Long Short-Term Memory (LSTM) model is well-suited for capturing such dependencies by maintaining a "memory" of relevant information while filtering out noise \cite{Fischer, Greg, Lei, Sak}. Unlike traditional statistical techniques that assume linearity, LSTM learns complex temporal patterns that may not be explicitly defined in financial formulas. In this study, we leverage LSTM not for prediction, but for dimensionality reduction, compressing financial signals into a single latent representation that preserves key financial dynamics over time.

\begin{figure}[H]
    \centering
\includegraphics[width=\textwidth]{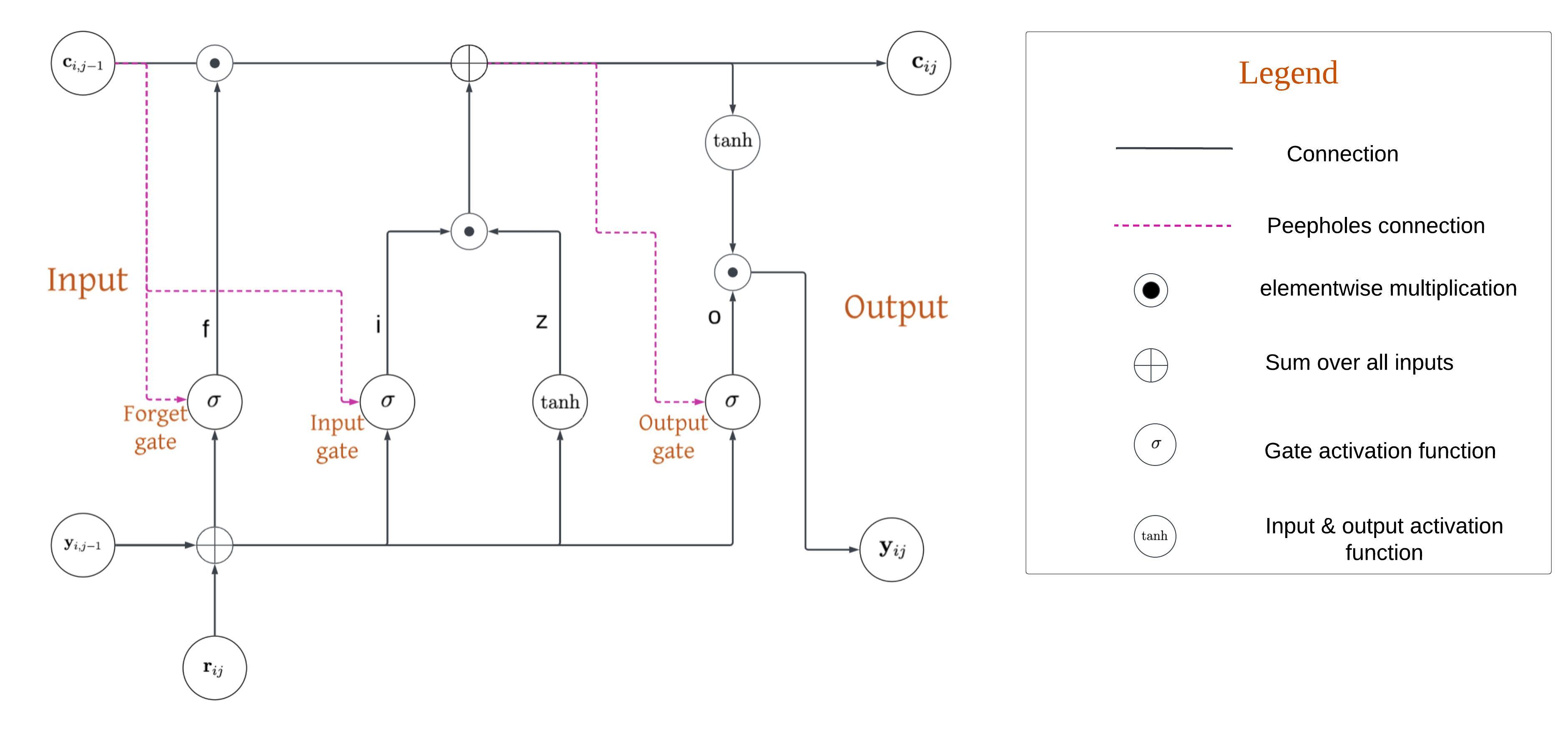}
\caption{\footnotesize Visualization of the architecture of a standard LSTM block, illustrating its key components, including the input gate, forget gate, and output gate, as well as the cell state and hidden state transitions. A legend is provided to clarify the flow of information and operations within the block.}
\label{fig: lstm}
    \end{figure}

The LSTM architecture is a specialized recurrent neural network (RNN) designed to capture long-range dependencies by addressing the vanishing gradient problem through gating mechanisms. Each LSTM memory block consists of a cell state \(\mathbf{c}_{i,j}\) that stores long-term information and three gates, input, forget, and output, that regulate information flow. Given an input sequence \(\mathbf{R}\in \mathbb{R}^{N \times J \times F}\) consisting of financial records from \(N \) insurance companies over \(J\) time points with \(F\) financial ratios per time step as introduced in Table \ref{tab: Ratios}. The goal is to reduce \(F\) to a single temporal dimension \(F'\), resulting in a compressed representation \(\mathbf{Z} \in \mathbb{R}^{N \times J \times F'}\).
The LSTM memory block operates as follows:
\begin{itemize}
    \item Block Input: The block input \(\mathbf{g}_{i,j} \in \mathbb{R}^y\) combines the current input \(\mathbf{r}_{i,j}\) and the previous hidden state \(\mathbf{y}_{i,j-1}\):
    \[
    \mathbf{g}_{i,j} = \tanh(\mathbf{W}_g \mathbf{r}_{i,j} + \mathbf{R}_g \mathbf{y}_{i,j-1} + \mathbf{b}_g),
    \]
    where \(\mathbf{W}_g \in \mathbb{R}^{y \times F}\), \(\mathbf{R}_g \in \mathbb{R}^{y \times y}\), and \(\mathbf{b}_g \in \mathbb{R}^y\) are learnable parameters.
    \item Input Gate: Determines the extent to which \(\mathbf{g}_{i,j}\) influences the cell state:
    \[
    \mathbf{i}_{i,j} = \sigma(\mathbf{W}_i \mathbf{r}_{i,j} + \mathbf{R}_i \mathbf{y}_{i,j-1} + \mathbf{p}_i \odot \mathbf{c}_{i,j-1} + \mathbf{b}_i),
    \]
    where \(\mathbf{W}_i \in \mathbb{R}^{y \times F}\), \(\mathbf{R}_i \in \mathbb{R}^{y \times y}\), \(\mathbf{b}_i \in \mathbb{R}^y\), and \(\mathbf{p}_i \in \mathbb{R}^y\) are learnable parameters.

    \item Forget Gate: Determines which information from the previous cell state is retained:
    \[
    \mathbf{f}_{i,j} = \sigma(\mathbf{W}_f \mathbf{r}_{i,j} + \mathbf{R}_f \mathbf{y}_{i,j-1} + \mathbf{p}_f \odot \mathbf{c}_{i,j-1} + \mathbf{b}_f),
    \]
    where \(\mathbf{W}_f \in \mathbb{R}^{y \times F}\), \(\mathbf{R}_f \in \mathbb{R}^{y \times y}\), \(\mathbf{b}_f \in \mathbb{R}^y\), and \(\mathbf{p}_f \in \mathbb{R}^y\) are learnable parameters.

    \item Cell State Update: Updates the cell state based on retained and new information:
    \[
    \mathbf{c}_{i,j} = \mathbf{f}_{i,j} \odot \mathbf{c}_{i,j-1} + \mathbf{i}_{i,j} \odot \mathbf{g}_{i,j}.
    \]

    \item Output Gate: Controls the information passed to the final output:
    \[
    \mathbf{o}_{i,j} = \sigma(\mathbf{W}_o \mathbf{r}_{i,j} + \mathbf{R}_o \mathbf{y}_{i,j-1} + \mathbf{p}_o \odot \mathbf{c}_{i,j} + \mathbf{b}_o),
    \]
    where \(\mathbf{W}_o \in \mathbb{R}^{y \times F}\), \(\mathbf{R}_o \in \mathbb{R}^{y \times y}\), \(\mathbf{b}_o \in \mathbb{R}^y\), and \(\mathbf{p}_o \in \mathbb{R}^y\) are learnable parameters.

    \item Hidden State Update: Computes the updated hidden state:
    \[
    \mathbf{y}_{i,j} = \tanh(\mathbf{c}_{i,j}) \odot \mathbf{o}_{i,j}.
    \]

    \item LSTM Output: A fully connected dense layer maps the hidden state to the output:
    \[
    \mathbf{z}_{i,j} = \mathbf{W}_z \mathbf{y}_{i,j} + \mathbf{b}_z,
    \]
    where \(\mathbf{W}_z \in \mathbb{R}^{1 \times y}\) and \(\mathbf{b}_z \in \mathbb{R}^1\) are learnable parameters.
\end{itemize}
The overall transformation performed by the LSTM network can be expressed as:
\[
\mathbf{Z} = \psi_\text{LSTM}(\mathbf{R}),
\]
where \(\psi_\text{LSTM}\) denotes the LSTM's mapping from the input space \(\mathbb{R}^{N \times J \times F}\) to the output space \(\mathbb{R}^{N \times J \times F'}\). Figure \ref{fig: lstm} displays the LSTM architecture. 
Unlike static dimensionality reduction methods such as PCA or t-SNE, which do not account for temporal dependencies, LSTM dynamically transforms financial data while preserving its sequential nature. By applying LSTM, we extract a latent representation that summarizes evolving financial trajectories while retaining meaningful variations. This compressed representation ensures that firms with similar financial evolution are mapped closer together in latent space, preserving patterns that static methods may overlook.

\begin{figure}[H]
\begin{center}
\includegraphics[width=\columnwidth]{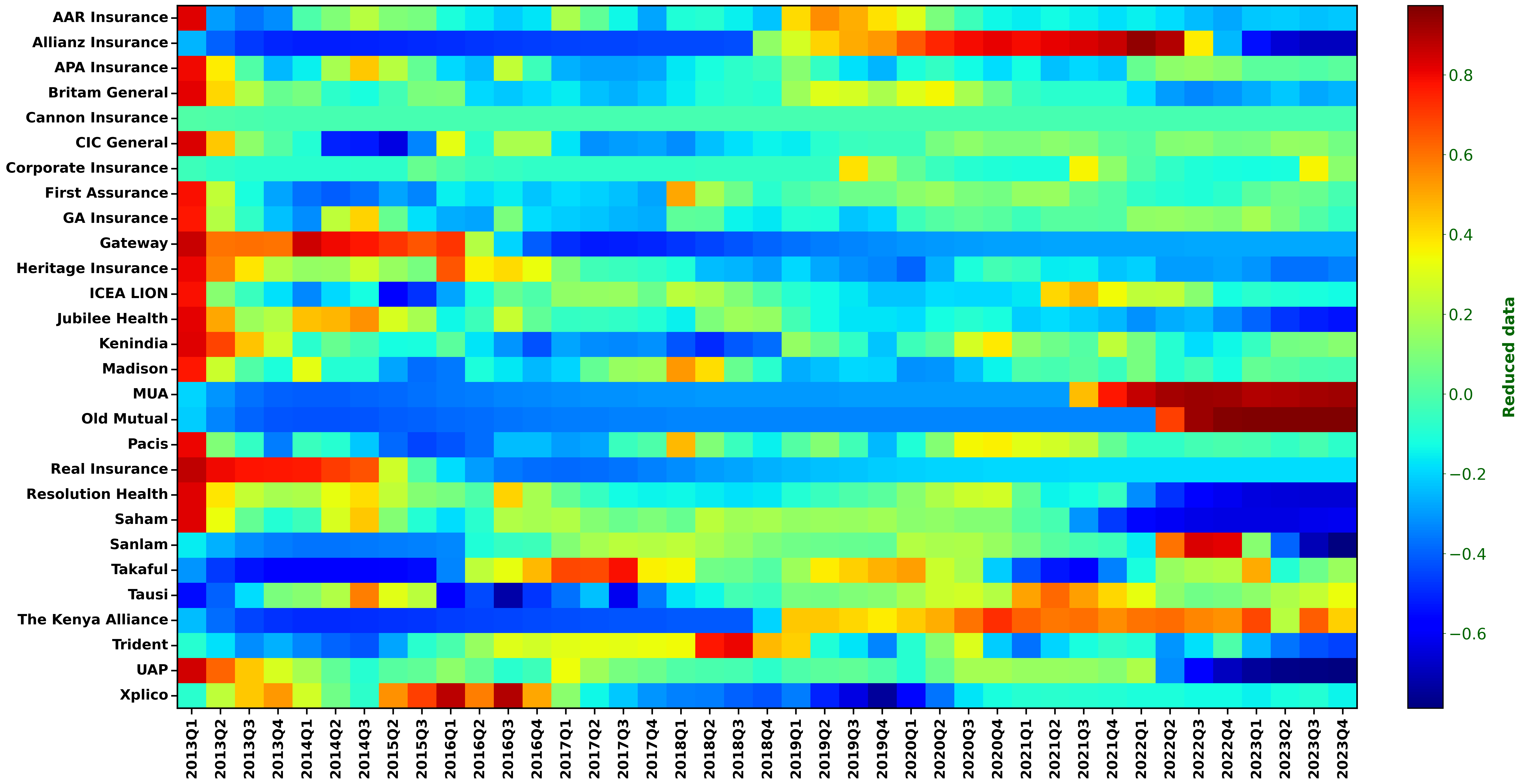}
\caption{\footnotesize The heatmap displays the reduced time series for each company after applying temporal feature fusion using LSTM, which transforms the data from \(\mathbb{R}^7\) to \(\mathbb{R}^1\). Color intensity indicates the magnitude of the reduced data values, with darker shades representing higher values. The x-axis represents time points, and the y-axis lists the companies, which will be clustered based on this dataset.}
\label{fig: lstm_heatmap}
\end{center}
\end{figure}

The LSTM model was configured with a single hidden layer consisting of \(y = 64\) units and trained for 12 epochs with a batch size of 16. The Adam optimizer minimized the mean squared error (MSE) loss function:
   \[
   \mathcal{L}_\text{LSTM}(\mathbf{R}, \hat{\mathbf{R}}) = \frac{1}{N \cdot J \cdot F} \sum_{i=1}^N \sum_{j=1}^J \sum_{f=1}^F \left( r_{i,j,f} - \hat{r}_{i,j,f} \right)^2,
   \]
where \(\hat{\mathbf{R}}\) is the reconstructed input tensor. The dataset used in this study consists of \textbf{28 insurance companies}, each observed over \textbf{41 time points} with \textbf{7 financial ratios} (as detailed in Table~\ref{tab: Ratios}). The LSTM model was applied to extract a lower-dimensional representation, reducing the input tensor 
\(\mathbf{R} \in \mathbb{R}^{28 \times 41 \times 7}\) to a more compact form, 
\(\mathbf{Z} \in \mathbb{R}^{28 \times 41 \times 1}\). 
This transformation preserves essential financial patterns while maintaining temporal dependencies. The extracted latent features, visualized in Fig.~\ref{fig: lstm_heatmap}, serve as the foundation for the clustering analysis discussed in Section~\ref{sec:clusters}.

\subsection{Time Series Clustering}\label{sec:clusters}
Clustering time-series data is crucial for identifying trends and uncovering patterns across multiple entities \cite{Evans}. In this study, we group insurance companies based on the latent representations $\mathbf{Z} \in \mathbb{R}^{N \times J \times 1}$ obtained from the LSTM model. These representations capture the temporal dynamics of financial ratios for each company while reducing the dimensionality of the data. Given the temporal nature of the data and potential structural variations, we employ the Dynamic Time Warping (DTW) algorithm \cite{Kim}, a robust method for measuring time-series similarity.
DTW is particularly suitable for datasets with temporal misalignments or missing values, as it aligns time-series data of unequal lengths \cite{Li}. Its insensitivity to outliers ensures that unusual data points do not disproportionately affect similarity measurements \cite{Pavel}, making DTW ideal for clustering time-series data characterized by temporal shifts or distortions.

To illustrate the DTW framework, consider the latent representations of two insurers as $\mathbf{z}_A = \{z_{A1}, z_{A2}, \dots, z_{AJ}\}$ and $\mathbf{z}_B = \{z_{B1}, z_{B2}, \dots, z_{BJ}\}$, where each $z_{Aj}, z_{Bj} \in \mathbb{R}^1$ represents the compressed financial data at time $j$. The DTW algorithm computes the similarity between these two time series by constructing a cost matrix $D$ of size $J \times J$. Each element $D(k, l)$ represents the squared Euclidean distance between the latent variables at time points $k$ and $l$, expressed as:
\[
D(k, l) = (z_{Ak} - z_{Bl})^2.
\]

To align the two time series, the algorithm identifies a warping path $W = \{(k_1, l_1), (k_2, l_2), \dots, (k_K, l_K)\}$, which maps the indices $k$ and $l$ to account for temporal differences. This warping path satisfies three key conditions: the boundary condition (starting at $(1,1)$ and ending at $(J, J)$), continuity (adjacent steps in $W$ are allowed), and monotonicity (indices in $W$ do not decrease).

The goal of DTW is to minimize the cumulative warping cost, defined as:
\[
\text{Cost}(W) = \frac{1}{K} \sum_{k=1}^{K} D(k_k, l_k),
\]
where $K$ is the length of the warping path. This minimization ensures that similar temporal patterns in the latent representations are effectively matched.

The optimal warping path is computed using dynamic programming \cite{Salvador, Keogh}. A cumulative distance matrix $\Gamma(k, l)$ is recursively calculated as:
\[
\Gamma(k, l) = D(k, l) + \min\{\Gamma(k-1, l-1), \Gamma(k-1, l), \Gamma(k, l-1)\},
\]
where $\Gamma(k, l)$ represents the minimum cumulative cost up to cell $(k, l)$. The optimal path is determined by backtracking from $\Gamma(J, J)$.

\begin{figure}[H] 
\begin{center} 
 \includegraphics[width=0.46\columnwidth]{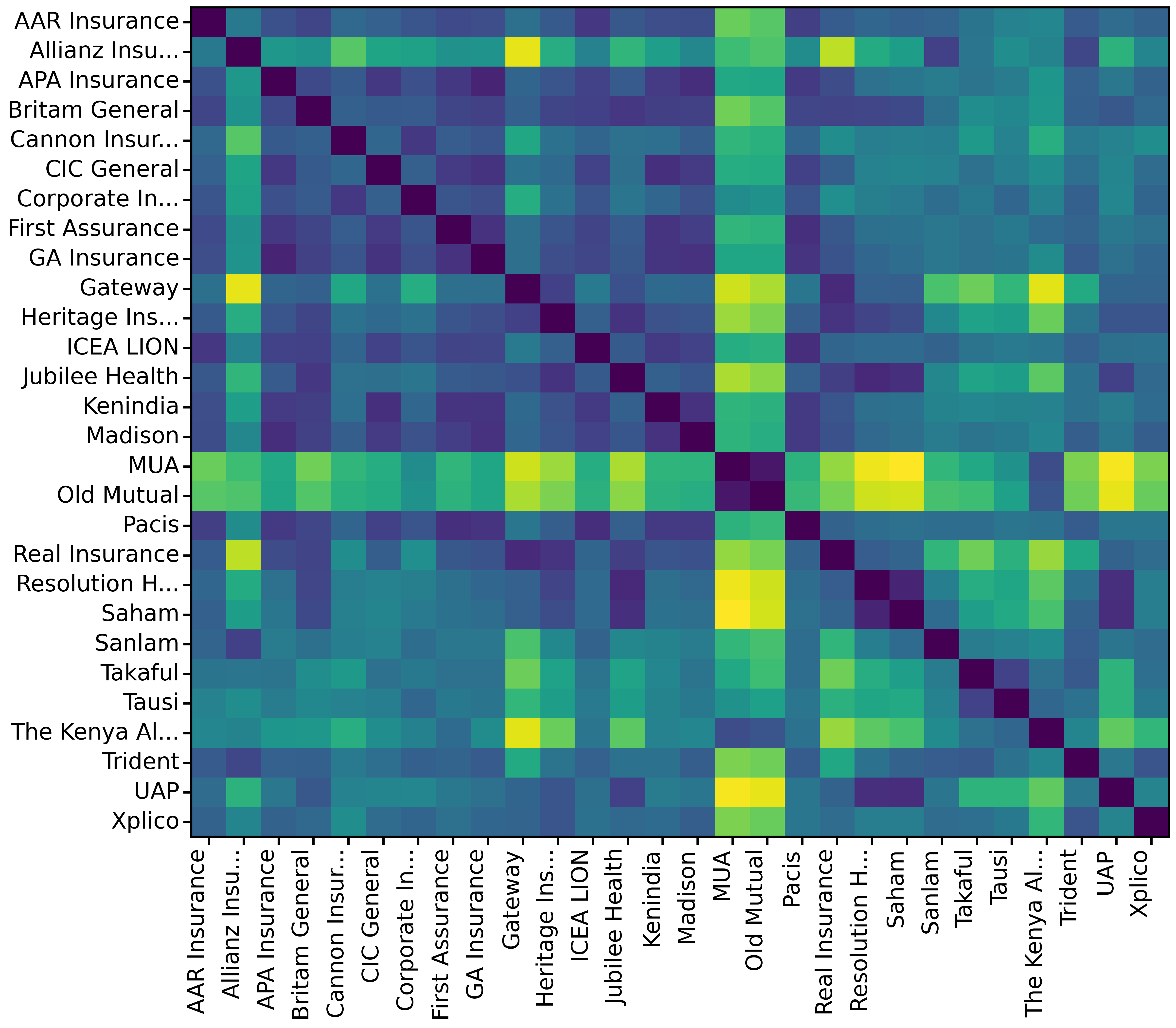}  \includegraphics[width=0.53\columnwidth]{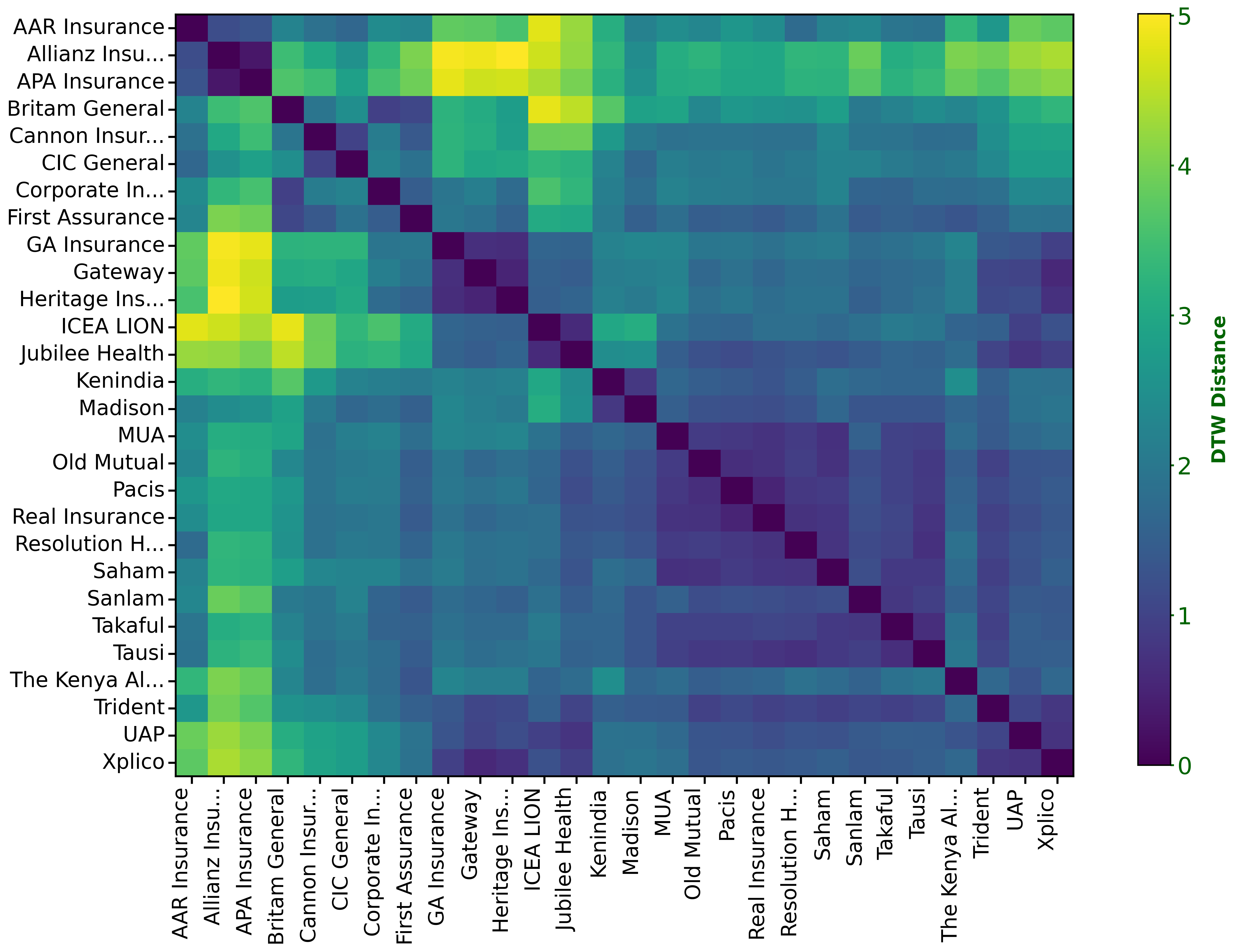} 
 \caption{
        \footnotesize
        On the left is the unordered DTW distance matrix, which visualizes the pairwise distances between time series for all insurance companies without any specific order. On the right is the ordered DTW distance matrix, where the rows and columns are rearranged based on hierarchical clustering (complete linkage method). This reordering groups companies with similar time series patterns, allowing for a clearer identification of relationships and clusters within the data. The color intensity represents the magnitude of the DTW distance, with darker colors indicating smaller distances.
    }
    \label{fig:dtw_distance_matrix}    

\end{center} 
\end{figure}

Applying DTW to the latent representations $\mathbf{Z}$ enables meaningful comparisons between companies despite temporal distortions. Pairwise DTW distances between the latent time series of all insurers are computed to construct a similarity matrix. This matrix serves as the input for clustering algorithms such as hierarchical clustering or K-Means. The left panel of Fig. \ref{fig:dtw_distance_matrix} presents this similarity matrix. To enhance interpretability, rows and columns are rearranged using hierarchical clustering with complete linkage, as shown in the right panel of Fig. \ref{fig:dtw_distance_matrix}. This visualization highlights possible clusters of companies with similar financial behavior over time \cite{Ev, Ko}.

\subsection{Evaluation of Clusters}
The quality of the clusters identified in this analysis is evaluated using two widely recognized metrics: the Silhouette Score and the Elbow Method. These approaches provide complementary insights into the cohesiveness and separability of clusters and help determine the optimal number of clusters.
The Silhouette Score measures how well-separated the clusters are by assessing the similarity of each latent representation \( z_{ij} \) to its assigned cluster compared to other clusters \cite{Dzuba}. For a given latent value \( z_{ij} \), the Silhouette Score \( s_{ij} \) is calculated as:
\[
s_{ij} = \frac{b_{ij} - a_{ij}}{\max(a_{ij}, b_{ij})},
\]
where \( a_{ij} \) represents the average distance from \( z_{ij} \) to all other points within the same cluster, and \( b_{ij} \) is the average distance from \( z_{ij} \) to all points in the nearest neighboring cluster. The score \( s_{ij} \) ranges from -1 to 1, where a value close to 1 indicates strong clustering, a value close to -1 suggests misclassification, and a value near 0 implies that the point is near a cluster boundary.

To evaluate overall clustering quality, the mean silhouette score \( \overline{s} \) is computed across all latent representations:
\[
\overline{s} = \frac{1}{N \times J} \sum_{i=1}^{N} \sum_{j=1}^{J} s_{ij},
\]
where \( N = 28 \) is the total number of companies and \( J = 41 \) is the number of time points. A higher mean silhouette score indicates better-defined clusters \cite{Albert, Peter}. The silhouette score is calculated using the pairwise distance matrices derived from the Dynamic Time Warping (DTW) analysis, ensuring that temporal alignments are considered in the evaluation.

The Elbow Method complements this analysis by identifying the optimal number of clusters based on the total within-cluster variance, or distortion, as a function of the number of clusters \cite{Agga}. The distortion \( v_m \) for a given number of clusters \( m \) is defined as:
\begin{equation}
v_m = \sum_{j=1}^{m} \sum_{z_{ij} \in C_j} \| z_{ij} - \mu_j \|^2,
    \label{eq:elbow}
\end{equation}
where \( C_j \) represents the set of latent values assigned to cluster \( j \), and \( \mu_j \) is the centroid of cluster \( j \) in the latent space. This measure quantifies cluster compactness, with lower values indicating tighter groupings. By plotting distortion against different values of \( m \), the optimal number of clusters is identified at the "elbow" point, where the rate of reduction in distortion significantly decreases. Beyond this point, adding more clusters provides minimal improvement in clustering quality.

\begin{figure}[H] 
\begin{center} 
    \includegraphics[width=0.48\columnwidth]{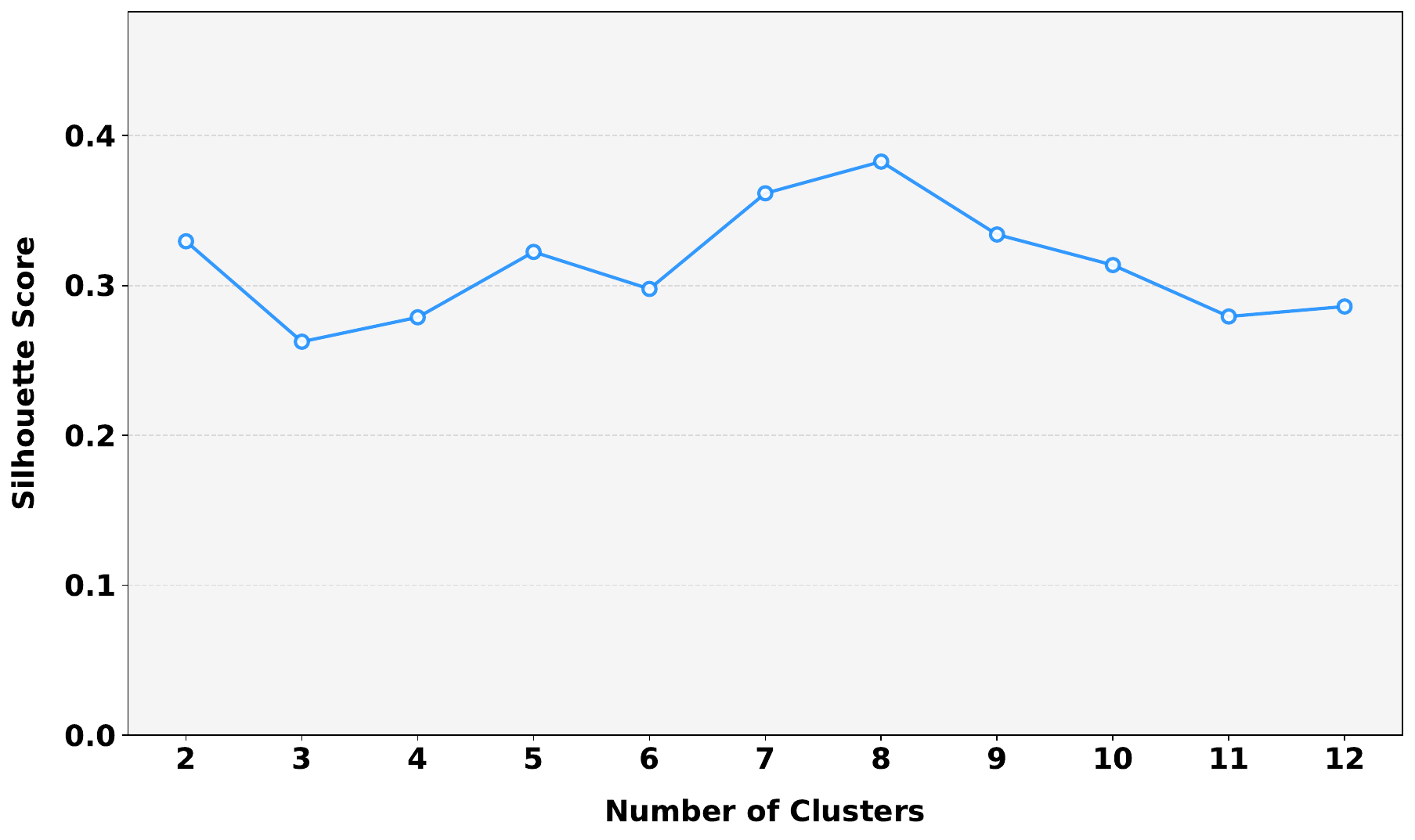}  
    \includegraphics[width=0.5\columnwidth]{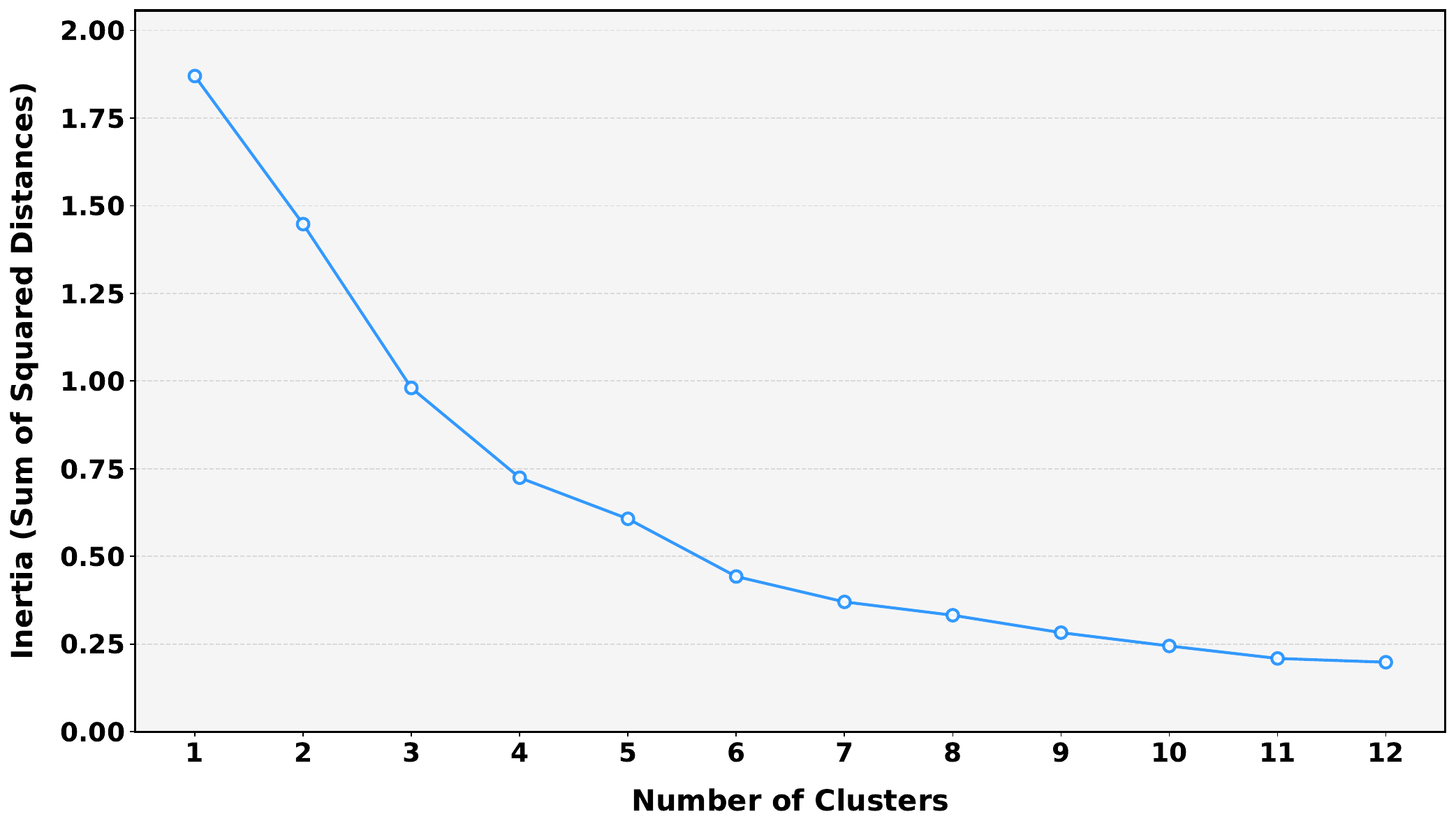}  
    \caption{\footnotesize The left plot shows the Silhouette Curve, indicating clustering quality, where higher values suggest better separation and cohesion of clusters. The right plot illustrates the Elbow Method, which identifies the optimal cluster number by the "elbow" point where the rate of decrease in inertia slows.}
    \label{fig:cluster_evaluation}
\end{center} 
\end{figure}

This study applies both the Silhouette Score and the Elbow Method to evaluate the DTW-based clustering results, ensuring a robust assessment. The outcomes of these evaluations are illustrated in Fig. \ref{fig:cluster_evaluation}, presenting a side-by-side comparison for comprehensive analysis.

\subsection{Time Series Financial Performance Analysis}
The time-series analysis of financial performance provides crucial insights into the trends and dynamics shaping the Kenyan medical insurance industry. The five heatmaps illustrate variations in key financial indicators, including market share, net earned premium income, underwriting profit, the number of new policies, and total policies issued over the years. The heatmaps present the trends across different insurance companies from 2013 to 2023.

\begin{figure}[H]
\begin{center}
    \includegraphics[width=\columnwidth]{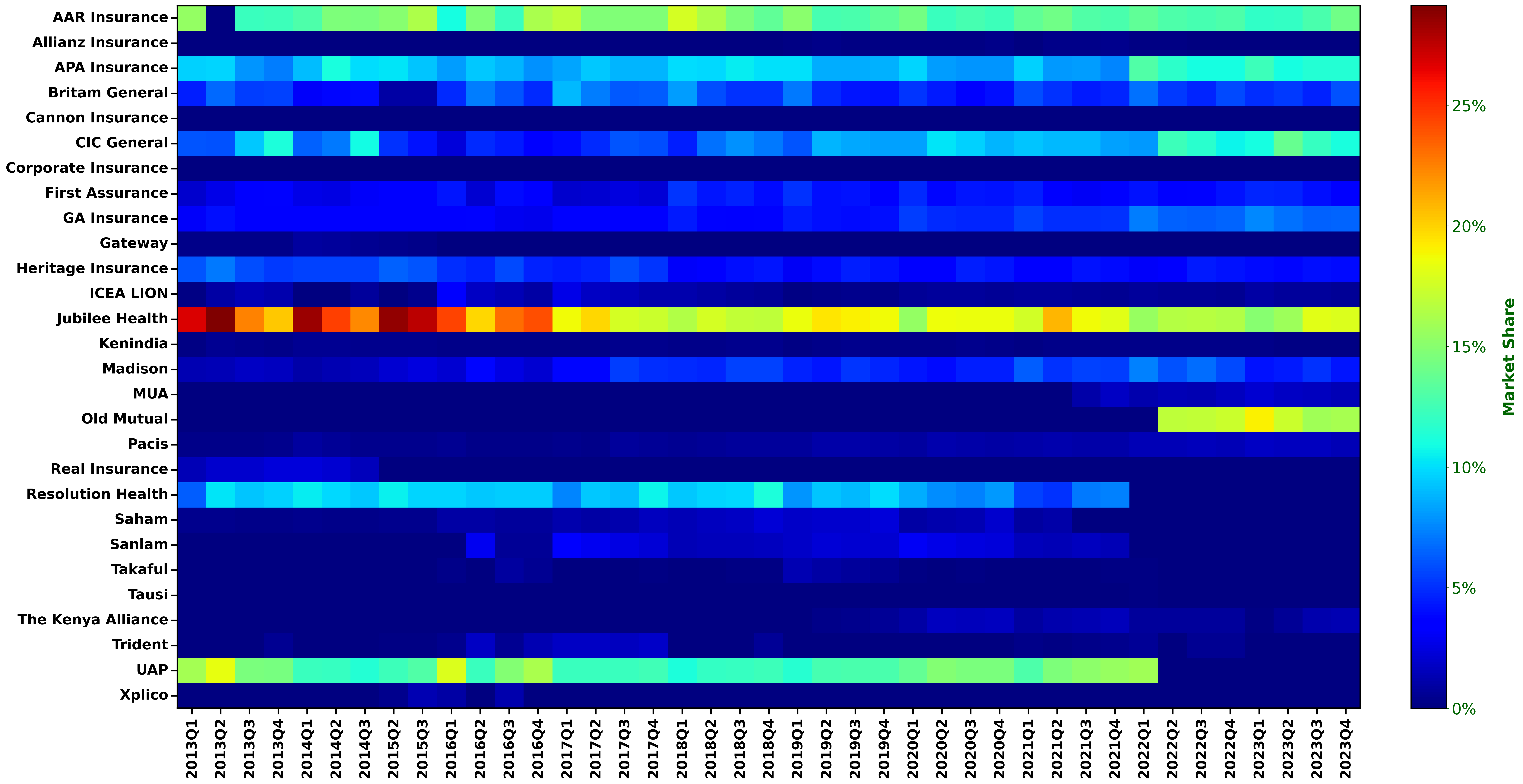}   
    \caption{\footnotesize Heatmaps illustrating the time series of the market share distribution among insurers. The color intensity represents the proportion of the total market share at each time period, with higher values indicated by red shades. At each time point, the market shares sum to 100\%.    
    }
    \label{fig: market share}
\end{center}
\end{figure}

\begin{figure}[H]
\begin{center}
    \includegraphics[width=0.49\columnwidth]{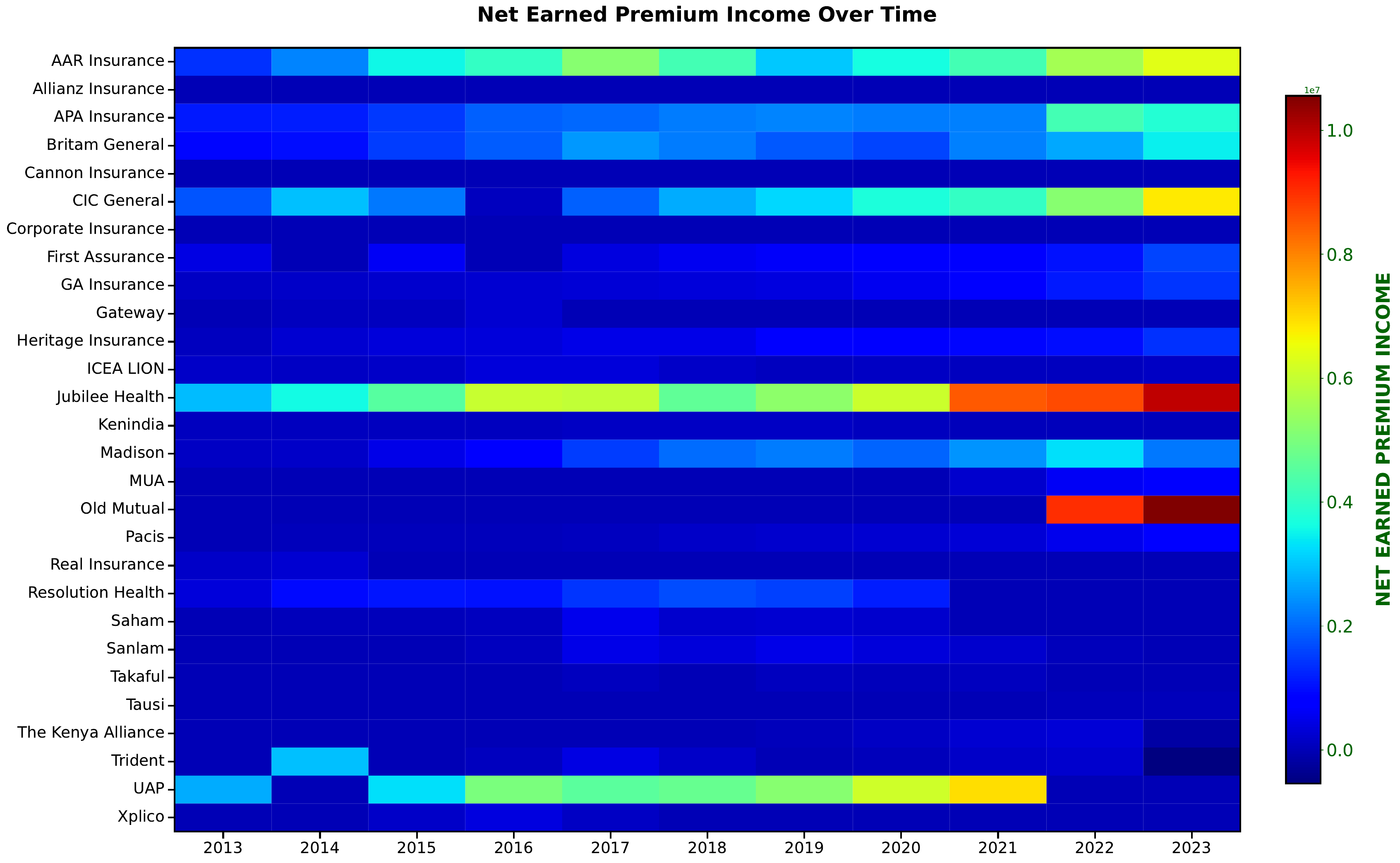}  
    \includegraphics[width=0.49\columnwidth]{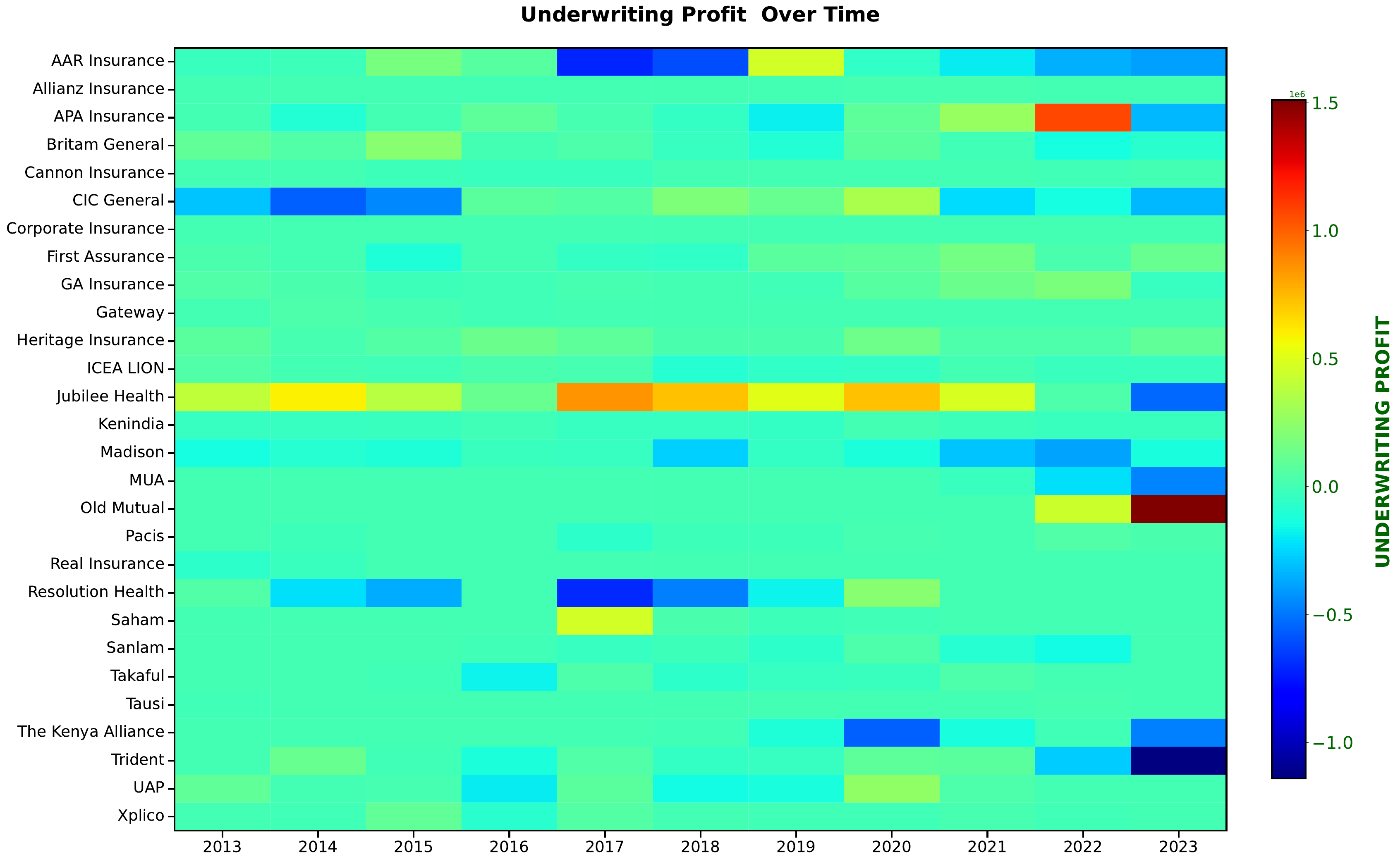} 
    
    \includegraphics[width=0.49\columnwidth]    {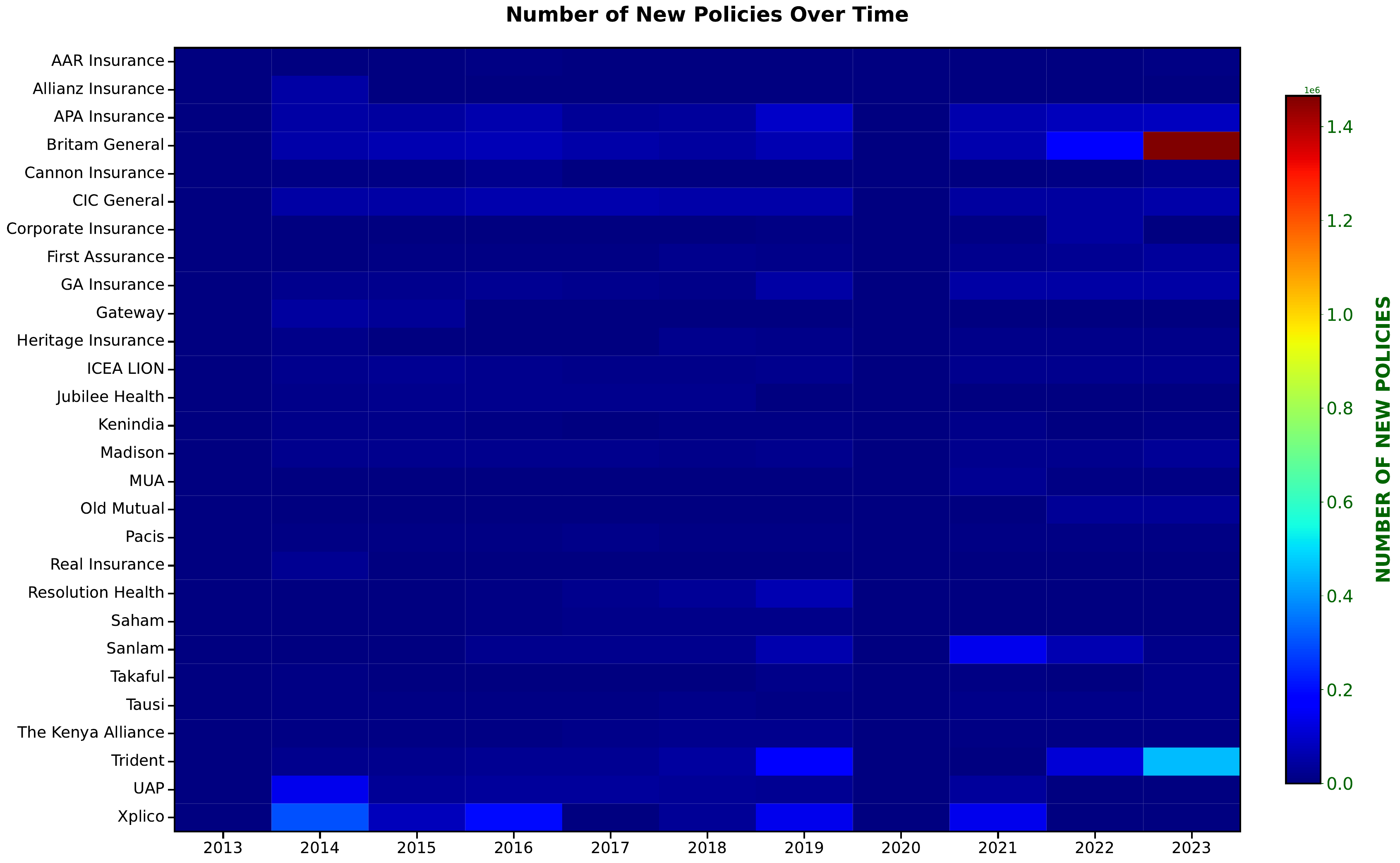}  
    \includegraphics[width=0.49\columnwidth]{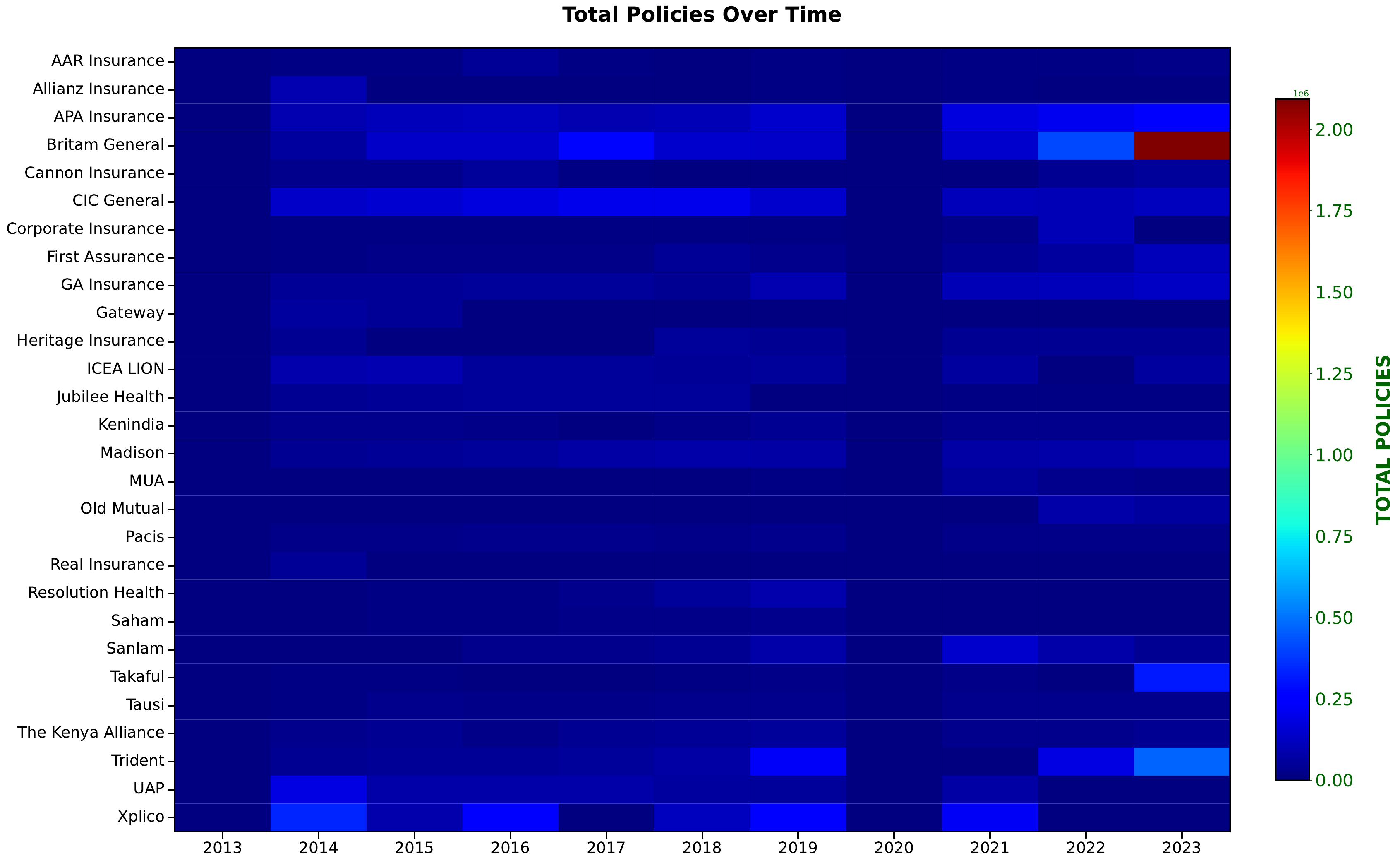} 
    \caption{\footnotesize Heatmaps illustrating key financial and operational metrics of various insurance companies over time. (Top Left) Net Earned Premium Income trends from 2013 to 2023. (Top Right) Underwriting Profit patterns over the same period. (Bottom Left) Number of New Policies issued each year for the general insurance. (Bottom Right) Total Policies in force over time. The color intensity represents the magnitude of each metric, with higher values indicated by red shades.}
    \label{fig: financial_perf}
\end{center}
\end{figure}

The heatmap \ref{fig: market share} illustrates the distribution of market share among various insurers over time. The visualization provides insights into the competitive landscape of the insurance industry, highlighting both dominant players and those with fluctuating or marginal market shares. 
One of the most notable observations is the consistent dominance of Jubilee Health, which maintains the highest market share across multiple time periods. Other significant players, such as AAR and APA Insurance also hold substantial market shares in different periods. These companies appear to have maintained their competitive positioning over time. In contrast, several insurers experience noticeable fluctuations in their market share. Resolution Health, UAP, and Old Mutual demonstrate periods of growth, where they hold a larger portion of the market, followed by stabilization or decline. This suggests that while some firms experience temporary surges, maintaining a dominant position in the insurance sector requires sustained strategic efforts. Additionally, some insurers, such as CIC General and Britam General, maintain relatively stable but moderate market shares, indicating steady performance without dramatic shifts.
Net earned premium income, shown in Figure \ref{fig: financial_perf} (top left), represents the portion of premiums retained after deductions for reinsurance and adjustments for policy cancellations. A closer examination of the heatmap reveals that while some companies consistently report stable or increasing premium income, others experience significant fluctuations. Companies such as Takaful and Tausi demonstrate low but steady growth in premium income, indicating robust underwriting practices and customer retention. Conversely, firms such as Madison Insurance exhibit sporadic premium inflows, suggesting challenges in policy sales, customer acquisition, or retention.

The underwriting profit heatmap, depicted in Figure \ref{fig: financial_perf} (top right), highlights profitability trends across insurance firms. Companies like Jubilee Health and Heritage Insurance consistently show strong underwriting profits, reflecting effective risk management and operational efficiency. However, some insurers, including CIC General and Resolution Health, report periods of negative underwriting profit, suggesting high claims payouts or excessive operational costs. Periodic losses for multiple companies emphasize industry-wide challenges in maintaining sustainable underwriting margins, likely due to fluctuating claims and expense ratios.

The number of new policies heatmap, illustrated in Figure \ref{fig: financial_perf} (bottom left), provides insights into policy issuance trends across different insurers. A significant observation is that Britam General records a substantial increase in new policies in recent years, positioning it as a leading player in customer acquisition. However, firms such as MUA, Takaful, and Pacis report minimal new policy sales, suggesting difficulties in attracting new customers. These disparities in new policy issuance highlight differences in market penetration strategies and product offerings among insurers.

The total policies heatmap, as shown in Figure \ref{fig: financial_perf} (bottom right), provides insights into the overall size of an insurer’s customer base. Britam General maintains a substantial policyholder base, reflecting strong customer retention and renewal strategies. Conversely, insurers like AAR Insurance, Cannon Insurance, and MUA show relatively low total policy counts, suggesting weaker customer retention or lower demand for their insurance products. The presence of abrupt changes in policy numbers for some firms, such as Xplico, could indicate strategic shifts, acquisitions, or product restructuring.

\section{Clustering Results}

The LSTM architecture effectively compressed the high-dimensional dataset from \(\mathbb{R}^{N \times J \times F}\) to a more concise representation in \(\mathbb{R}^{N \times J \times F^\prime}\), preserving key temporal patterns essential for clustering and interpretation. By reducing the original 7-dimensional financial data vectors \(\mathbf{r}_{ij} \in \mathbb{R}^7\) to 1-dimensional latent representations \(z_{ij} \in \mathbb{R}^1\), the LSTM structure successfully captured meaningful temporal dependencies in the financial ratios of companies over time.
To improve interpretability, the financial ratios detailed in Table \ref{tab: Ratios} were normalized prior to dimensionality reduction. Pairwise Dynamic Time Warping (DTW) distances were then computed between the latent representations \(\mathbf{Z}\) of insurance companies and visualized as a distance matrix (Fig. \ref{fig:dtw_distance_matrix}, left). This matrix was reorganized using a dendrogram to enhance clarity, revealing the underlying cluster structures (Fig. \ref{fig:dtw_distance_matrix}, right). In the matrix, smaller blocks indicate tightly clustered companies with high similarity, while larger blocks reflect groups with greater dissimilarity. For instance, the darker regions corresponding to MUA and Old Mutual form a distinct cluster, highlighting their strong similarity as shown in the reorganized matrix.

Determining the optimal number of clusters, \(m\), was a critical step in the analysis. The silhouette score and the elbow method guided this decision (see Fig. \ref{fig:cluster_evaluation}). The silhouette score measures how well-separated the clusters are and indicated diminishing improvements beyond \(m > 12\). The elbow method (Eq. \ref{eq:elbow}), which evaluates total within-cluster variance (Fig \ref{fig:cluster_evaluation}, right panel), suggested an optimal range of \(m\) between four and eight. For \(m = 8\), the silhouette score peaked at 0.35, though the resulting small clusters were difficult to interpret. This observation mirrors patterns in datasets like the Fisher Iris dataset, where \(m = 2\) yields the highest silhouette score despite the presence of three natural clusters \cite{Fisher}. To address this, both methods were combined for a more nuanced decision. Consequently, \(m = 4\) was chosen, yielding a mean silhouette score of 0.26 and aligning with the elbow point. This choice balances interpretability with clustering precision, facilitating practical analysis of the dataset's internal structure.

\subsection{K-Means Clustering}
\begin{figure}[H]
\begin{center}
    \includegraphics[width=0.49\columnwidth]{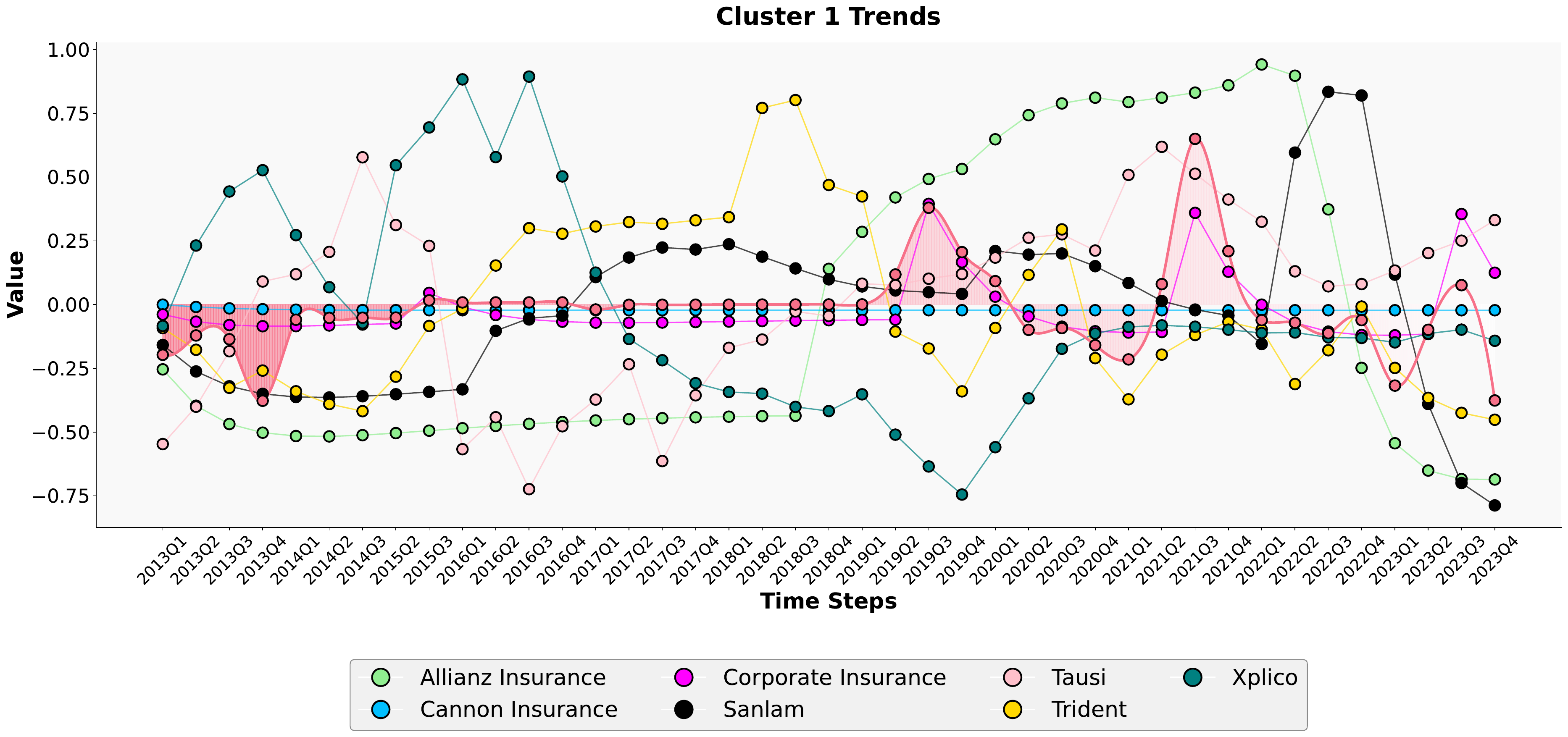}  
    \includegraphics[width=0.49\columnwidth]{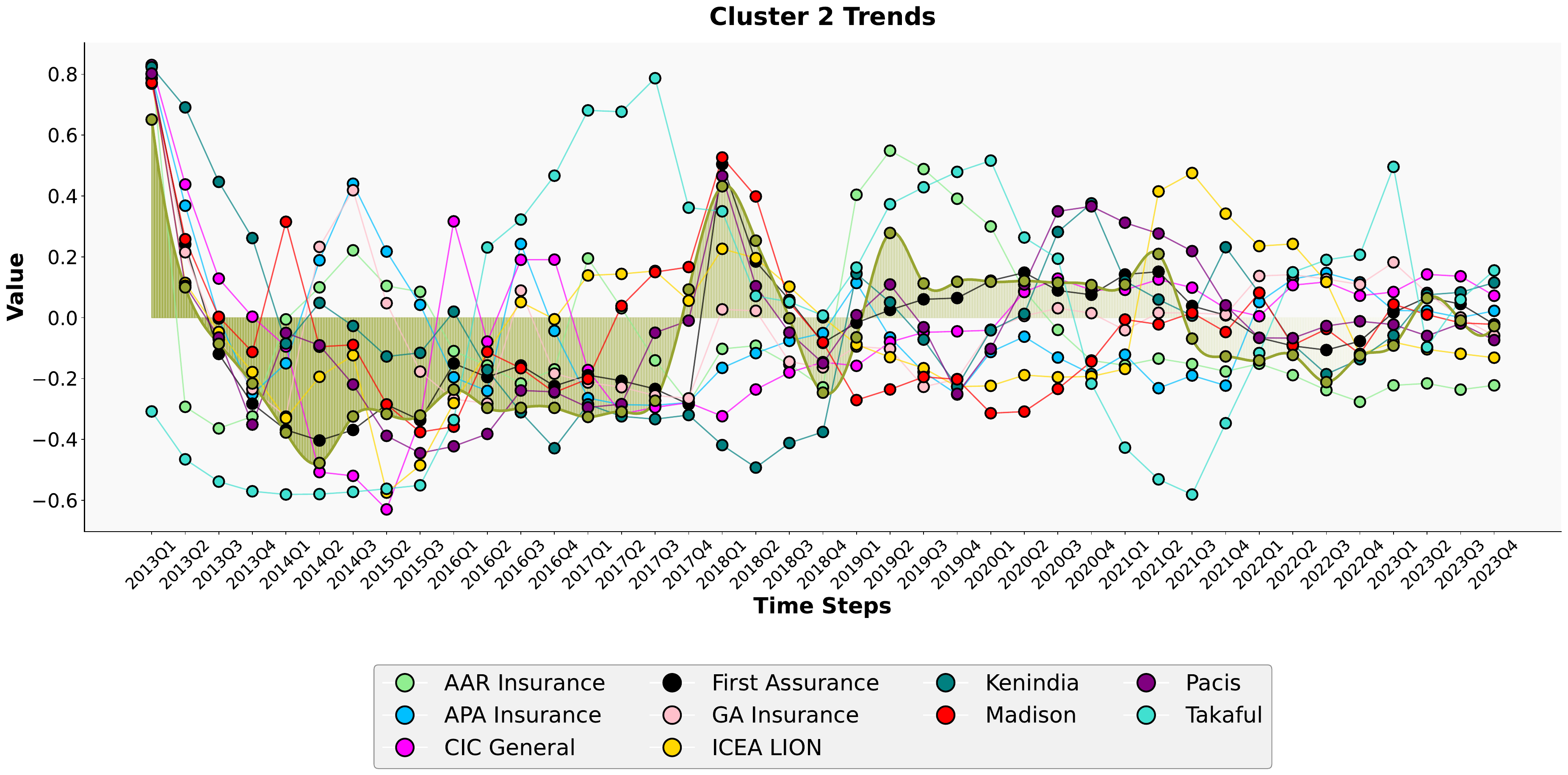}  
    
    \includegraphics[width=0.48\columnwidth]{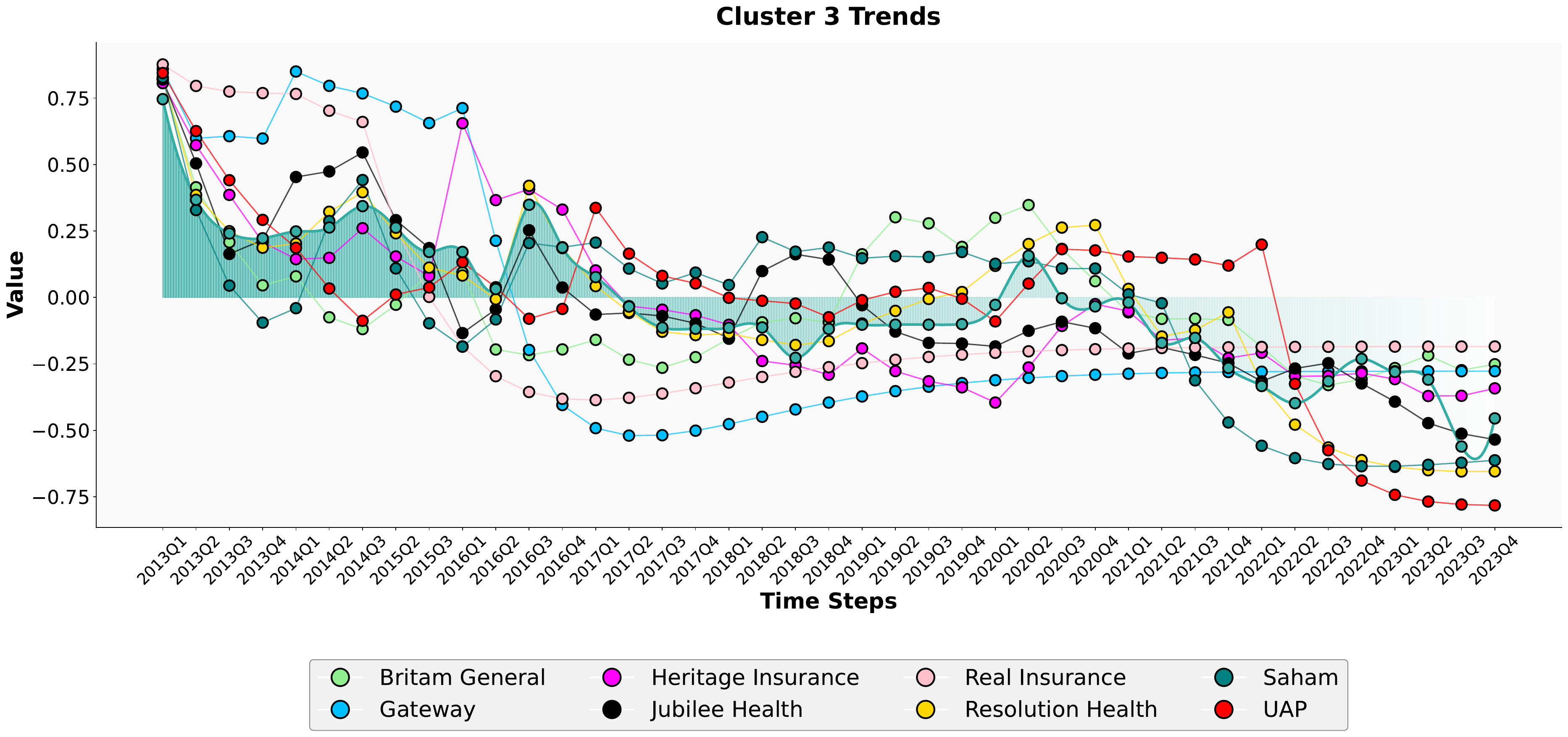}  
    \includegraphics[width=0.5\columnwidth]{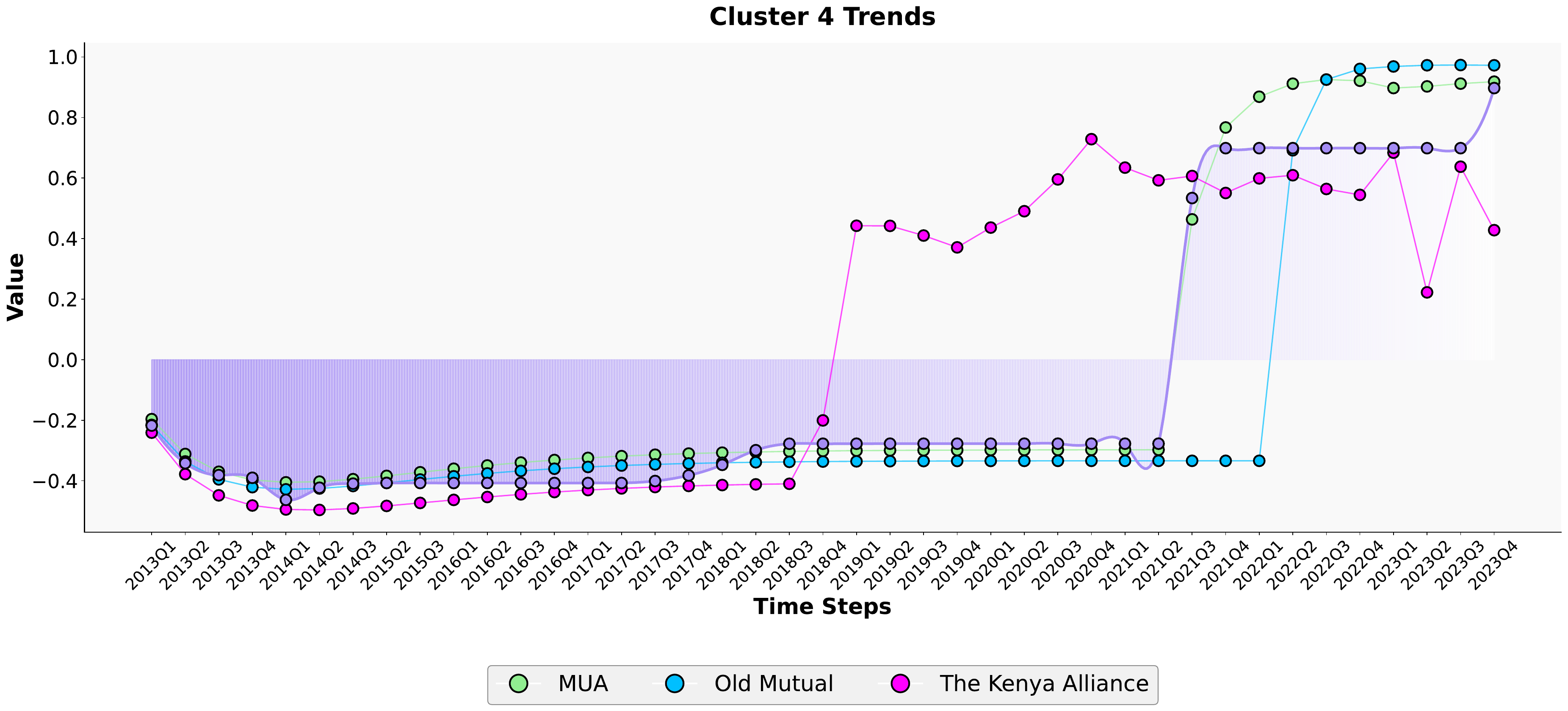}
    \caption{\footnotesize The figure presents scatter plots of the scaled latent time series for the four identified clusters. The top-left and top-right plots correspond to Clusters 1 and 2, respectively, while the bottom-left and bottom-right plots depict Clusters 3 and 4. Each company’s time series is color-coded within its respective cluster, with the bold lines representing cluster centers, which indicate the average latent time series trend. The gradient fill emphasizes overall cluster trends, while company names are annotated in the legend for easy identification. The x-axis represents time steps, and the y-axis denotes the first reduced feature values for each company. Cluster compositions include seven companies in Cluster 1, ten in Cluster 2, eight in Cluster 3, and three in Cluster 4.}
    \label{fig:time_series_clusters}
\end{center}
\end{figure}

K-Means clustering is a widely used unsupervised machine learning technique that partitions data into $m$ distinct clusters based on similarity measures. In this study, the \textit{TimeSeriesK-Means} algorithm was applied to group insurance companies according to the temporal patterns in their financial performance. Unlike traditional K-Means, which relies on Euclidean distance, \textit{TimeSeriesK-Means} incorporates Dynamic Time Warping (DTW) to account for temporal misalignments in financial data. 
In this study, we calculated Pairwise DTW distances between the latent time series, and the \textit{TimeSeriesK-Means} algorithm grouped companies accordingly. The clustering results for \(m = 4\) are presented in Fig. \ref{fig:time_series_clusters}, where the vertical axis represents the scaled latent time series data \(\mathbf{Z}\) across 41 quarters for each company. 
Cluster 1 comprises insurance companies that exhibit inconsistent financial performance, as reflected in the time series trends across various financial indicators. These companies show frequent fluctuations in their values, with periods of both high and low performance, as well as occasional gaps in their Gross Premium Income data. The cluster highlights erratic movements for several insurers, characterized by extreme peaks and troughs, indicating instability in their premium collections and overall financial performance. The presence of missing data points further underscores the possibility of financial distress or reporting inconsistencies, which may require regulatory intervention or targeted support. These companies generally hold low and fluctuating market shares, reinforcing their lack of stable positioning within the industry (see Fig. \ref{fig: financial_perf}, top left). Moreover, in the Net Earned Premium Income heatmap, some of these insurers display inconsistent earnings, with sporadic bright patches indicating occasional financial improvements. However, these improvements do not appear to be sustained over time, aligning with underwriting profit trends that also suggest volatility. The underwriting profit heatmap reveals that some of these firms experience periods of negative profits, signaling poor risk management or high claims ratios that impact financial stability. 
Interestingly, the new policies and total policies heatmaps provide additional insights into the performance of these firms. Many of these companies exhibit low or inconsistent new policy issuance, indicating weak customer acquisition strategies or an inability to expand their policyholder base. Some firms may experience occasional spikes in new policies, but these do not translate into sustained growth in total policies over time (see Fig. \ref{fig: financial_perf}). As shown in Fig. \ref{fig: market share}, these companies also struggle to maintain a consistent market share, which may indicate challenges related to customer acquisition, retention, or competitive pricing strategies.

Cluster 2 consists of insurers with relatively stable financial performance, punctuated by occasional fluctuations in profitability (Fig. \ref{sec:clusters}, top right). These companies generally exhibit strong financial resilience, managing to maintain consistent market positions despite periodic declines. The cluster includes a group of insurers such as AAR Insurance, APA Insurance, CIC General, ICEA LION, First Assurance, GA Insurance, Kenindia, Madison, Pacis, and Takaful, which navigate financial cycles with intermittent periods of decline. However, compared to firms in Cluster 1, these insurers demonstrate stronger recoveries and a general tendency to stabilize after downturns. The fluctuations observed in this cluster are less extreme than those in Cluster 1, indicating a more structured approach to financial and operational management. 
The market share heatmap (Fig. \ref{fig: market share}) reveals that many Cluster 2 insurers hold moderate to significant portions of the market share, reflecting their relatively stable competitive positioning. Companies such as ICEA LION and APA Insurance maintain consistent market shares, while others, like CIC General and First Assurance, show steady but moderate market penetration. Their ability to sustain their positions over time suggests strong customer trust, effective underwriting strategies, and solid financial management. Further insights can be drawn from the Net Earned Premium Income and Underwriting Profit heatmaps, where these firms demonstrate better financial stability than Cluster 1 companies. While sporadic downturns are observed, they do not persist for extended periods, suggesting that these firms have diversified revenue streams, efficient claims management, or effective cost-control measures that mitigate prolonged losses. The new policies and total policies heatmaps further support this classification, showing consistent issuance of new policies for most companies in this cluster. The ability to maintain a steady flow of new policies indicates market confidence and customer retention strategies that contribute to their financial resilience (see Fig. \ref{fig: financial_perf}).

Cluster 3 comprises companies that were once dominant players in the insurance industry but have experienced a transition in their financial performance over time (see Fig. \ref{sec:clusters}, bottom left). Contrary to earlier assumptions of data discontinuity, the time series trends, market share, and financial indicators reveal that these companies remain active, though their market positioning has evolved. Notably, firms such as Jubilee Health, Britam General, UAP, and Resolution Health have maintained substantial market presence and financial strength, even though their trajectories have fluctuated over time. 
The market share heatmap (Fig. \ref{fig: market share}) confirms this trend, showing that Jubilee Health, Britam General, and UAP have consistently held significant portions of the market over time. While their market shares may have fluctuated, they remain among the industry leaders, demonstrating resilience amid competitive pressures and economic shifts. Some companies in this cluster, such as Resolution Health, initially exhibited strong performance but later experienced gradual market share reductions, indicating possible strategic shifts, increased competition, or structural changes within the industry. Furthermore, the total policies and new policies heatmaps indicate that these insurers have maintained a steady flow of policy acquisitions over time. The Net Earned Premium Income heatmap shows that Jubilee Health, Britam General, and UAP continue to generate high premium revenues, reflecting strong underwriting performance and customer retention. The underwriting profit heatmap highlights some variability, with periods of high profitability followed by occasional dips. This suggests that while these companies have managed risk effectively, external shocks, regulatory changes, or claims volatility have impacted profitability at certain intervals (see Fig. \ref{fig: financial_perf}).

Cluster 4 (Fig. \ref{sec:clusters}, bottom right) represents a group of insurers that only began reporting financial data in later years but have since demonstrated strong financial performance and market expansion. These companies, including MUA, Old Mutual, and The Kenya Alliance, initially showed minimal or no data activity but have experienced rapid growth, outperforming other clusters in certain financial indicators. The market share heatmap (Fig. \ref{fig: market share}) supports this observation, as MUA, Old Mutual, and The Kenya Alliance show significant increases in market share in recent years. Unlike insurers in Cluster 3, which have gradually lost market share over time, these firms have expanded aggressively, capturing a larger portion of the industry. 
Additionally, the total policies and new policies heatmaps highlight a surge in policy issuance, confirming that these companies are actively acquiring new customers. Financially, Cluster 4 companies exhibit progressive improvements in premium income and underwriting profits. The Net Earned Premium Income heatmap shows a steady increase, indicating that these firms have successfully attracted policyholders and generated significant revenue from premiums. The underwriting profit heatmap reveals that their profitability has grown alongside premium earnings, suggesting effective risk management and claims control mechanisms (see Fig. \ref{fig: financial_perf}).

\subsection{Hierarchical Clustering}
\begin{figure}[H]
\begin{center}
    \includegraphics[width=\columnwidth]{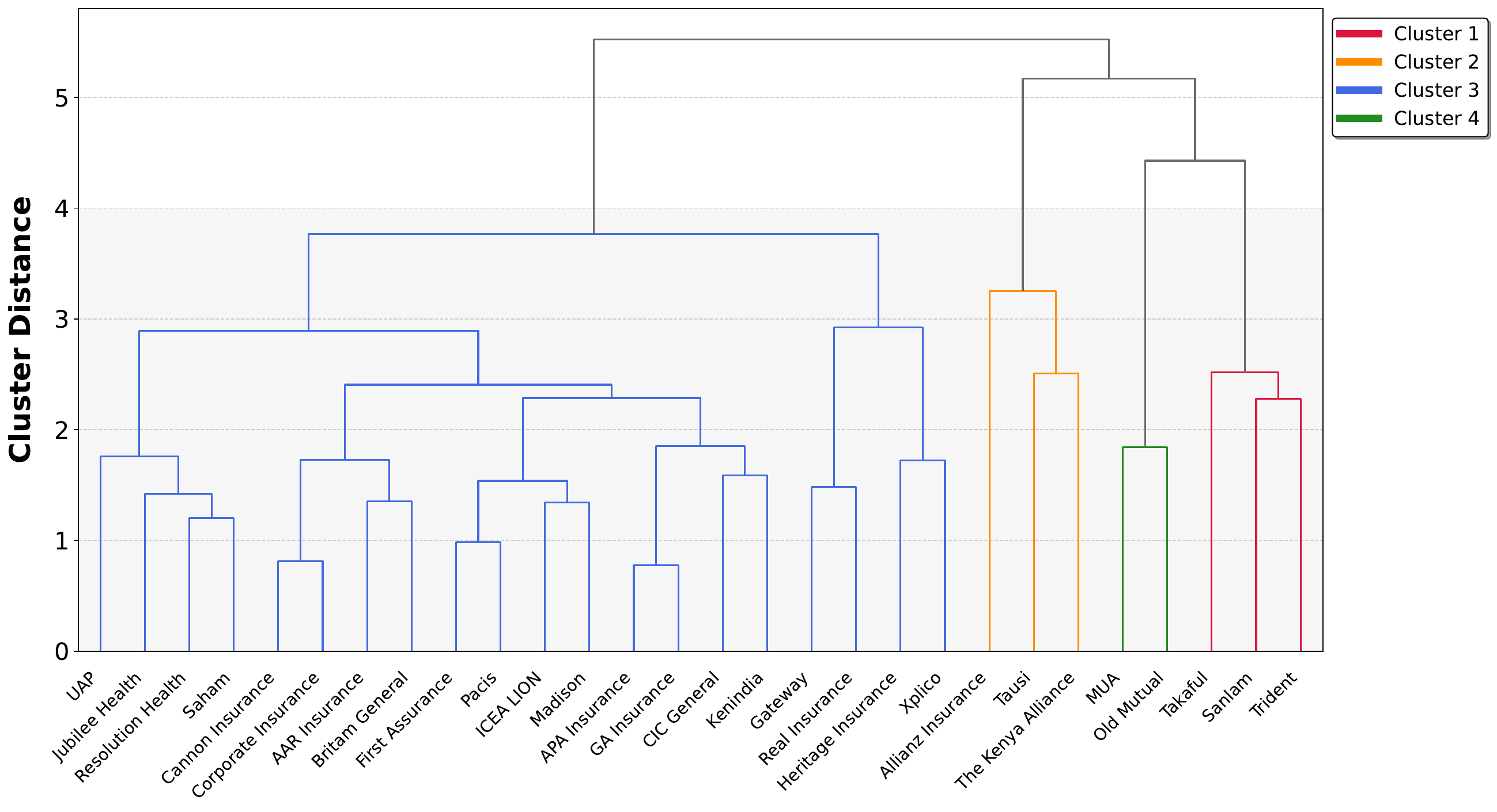} 
    \caption{\footnotesize Hierarchical clustering dendrogram of insurance companies based on their financial time series representations. The vertical axis represents cluster distance, illustrating the hierarchical relationships among companies. The four main clusters identified in the analysis are distinguished by different colors: Cluster 1 with three companies, Cluster 2 with three companies, Cluster 3 with eighteen companies, and Cluster 4 with two companies.}
    \label{fig: hier_cluster}
\end{center}
\end{figure}

Hierarchical clustering was performed to analyze the structural relationships between insurance companies based on their financial time-series data. Unlike K-Means, which partitions the data into a predefined number of clusters, hierarchical clustering organizes companies into a hierarchical structure, allowing for a more flexible and interpretable grouping process \cite{Evans}. 
The hierarchical clustering method employed a precomputed Dynamic Time Warping (DTW) distance matrix to capture temporal dependencies in financial ratios. Using the \textit{complete} linkage method, which minimizes the maximum inter-cluster distance during merging, the resulting dendrogram (Fig. \ref{fig: hier_cluster}) reveals well-separated groups with companies exhibiting high intra-cluster similarity.

A key advantage of hierarchical clustering is its ability to validate the strong relationships between closely connected companies. Unlike K-Means, which can be sensitive to initialization and predefined cluster counts, hierarchical clustering naturally groups companies with highly similar financial trajectories, as evidenced by the tightly clustered firms in the dendrogram. 
The dendrogram’s vertical axis represents cluster distances, illustrating which companies are most similar and which groups maintain a degree of separation. Firms that merge at lower distance thresholds exhibit near-identical financial behavior, making hierarchical clustering a valuable tool for confirming firm-level similarities with higher granularity than K-Means. For example, MUA and Old Mutual form a distinct, compact subgroup, reinforcing their strong financial resemblance. Similarly, Cannon Insurance and Corporate Insurance in Cluster 3 remain closely linked, confirming their shared market dynamics and financial stability over time.
The four-cluster partitioning aligns well with K-Means results while offering a more nuanced perspective on firm relationships. The hierarchical approach validates the grouping of insurance companies, reinforcing the robustness of the clustering methodology while allowing for a deeper exploration of industry dynamics.
The distribution of companies within these clusters identified by hierarchical clustering and K-Means is detailed in Table \ref{table: clusters}.
\begin{table}[H]
\centering
\begin{tabular}{||p{2.5cm}|p{6.2cm}|p{6.2cm}||}
\hline
\rowcolor{lightgray}
CLUSTER & K-Means Clusters & Hierarchical Clusters \\
\hline\hline
\rowcolor{gray!10}
1 & Allianz Insurance, Cannon Insurance, Corporate Insurance, Sanlam, Takaful, Tausi, Trident & Takaful, Sanlam, Trident \\
\hline
2 & AAR Insurance Kenya, APA Insurance, CIC General, First Assurance, GA Insurance, ICEA LION, Kenindia, Madison, Pacis, Xplico & Xplico, Allianz Insurance, Tausi, The Kenya Alliance \\
\hline
3 & Britam General, Gateway, Heritage Insurance, Jubilee Health, Real Insurance, Resolution Health, Saham, UAP & UAP, Jubilee Health, Resolution Health, Saham, Cannon Insurance, Corporate Insurance, AAR Insurance, Britam General, First Assurance, Pacis, ICEA LION, Madison, APA Insurance, GA Insurance, CIC General, Kenindia, Gateway, Real Insurance, Heritage Insurance \\
\hline
4 & MUA, Old Mutual, The Kenya Alliance & MUA, Old Mutual \\
\hline\hline
\end{tabular}
\caption{\footnotesize Comparison of insurance company assignments between K-Means and Hierarchical Clustering based on financial time series data. The table highlights similarities and differences in how each method grouped companies, emphasizing the closer relationships captured by hierarchical clustering.}
\label{table: clusters}
\end{table}

\section{Discussion}
The clustering results derived from financial ratios and time series data provide significant insights into the operational and financial behaviors of insurance companies. Identifying four distinct clusters enables a nuanced understanding of variations in performance, reporting consistency, and systemic challenges within the industry. The results indicate that certain insurers maintain financial stability and steady growth, while others exhibit volatility or inconsistent reporting practices, posing challenges to risk assessment and market transparency.

The implications of these findings extend to regulatory authorities, particularly the Kenya Insurance Regulatory Authority (IRA). Given the inconsistencies observed in Clusters 3 and 4, regulatory interventions could be implemented to enforce more stringent reporting standards. Mandatory reporting frameworks would ensure that all insurance companies submit timely and accurate financial data, reducing discrepancies and improving market transparency. Standardizing financial disclosures, particularly in underwriting profit and premium income, would allow for better comparability across insurers, facilitating more informed decision-making for both regulators and policyholders. Inconsistent reporting not only undermines risk assessment but also distorts industry performance metrics, which could lead to misinformed policy interventions. Enhanced regulatory oversight would mitigate these issues, ensuring that the insurance sector remains robust and accountable.

The clustering results also provide valuable insights for investors and policyholders in making informed decisions. A key takeaway from the analysis is that insurers can be categorized based on financial stability, growth trends, and risk exposure. Introducing a stability rating system could significantly enhance consumer confidence by classifying insurers based on underwriting profitability, policy retention rates, and premium consistency. Such a classification system would help policyholders identify insurers that are financially stable versus those exhibiting volatility, enabling better selection of insurance providers. Investors, on the other hand, could use these insights to assess risk-adjusted returns and make strategic investment decisions. This approach mirrors existing credit rating frameworks and could be adopted within the insurance sector to improve market efficiency. The classification of the companies aids stakeholders in identifying stable and high-risk insurers, making informed market decisions, and formulating risk mitigation strategies.

To ensure the robustness of the clustering methodology, hierarchical clustering was introduced alongside K-Means. While the clustering assignments remained largely consistent, hierarchical clustering provided further validation of the time-series-based groupings. Future work could incorporate additional cross-validation techniques, such as bootstrapping or reassignment consistency tests, to further confirm cluster stability. The application of Dynamic Time Warping (DTW) in this study is particularly suited to capturing temporal financial patterns, as it preserves the sequential dependencies in the data, distinguishing insurers with consistent long-term performance from those exhibiting sporadic fluctuations.

To enhance the contextual understanding of the clustered companies, additional firm-level attributes were considered, including market share, total policyholder base, and financial strength indicators such as underwriting profit trends. These attributes help explain variations in clustering results, as larger firms with significant market share tend to exhibit stable financial trends, whereas smaller firms or those with irregular policy issuances are more prone to volatility. The inclusion of this data provides a more holistic view of each insurer’s market positioning.
Macroeconomic variables, such as inflation, interest rates, and regulatory changes, can significantly impact an insurer’s financial performance. For instance, rising inflation can increase claims costs, thereby affecting underwriting profits. Similarly, changes in regulatory requirements, such as solvency mandates, can influence capital adequacy ratios, impacting insurer clustering. The results indicate that insurers in different clusters may respond differently to economic shocks; for example, companies in Cluster 2 may exhibit resilience, whereas those in Cluster 1 may face greater financial distress. Future research could integrate macroeconomic indicators to further analyze these relationships.

This study contributes to both theory and practice by leveraging DTW-based clustering, which offers a novel approach to analyzing financial time-series data within the insurance sector. Compared to conventional clustering methods such as PCA or Gaussian Mixture Models, DTW captures temporal dependencies more effectively, leading to more accurate segmentation of insurers. These insights align with previous studies that emphasize the role of time-series analysis in financial modeling. From a managerial perspective, the results provide actionable insights. Companies in Cluster 1 could benefit from improving operational efficiency and risk management to stabilize their financial performance. Regulators could prioritize firms in Clusters 3 and 4, enforcing stricter reporting requirements to enhance data consistency. Insurers within Cluster 2 can leverage their financial stability to expand market share and introduce innovative insurance products.

Despite this analysis's strengths, several limitations must be acknowledged. 
The analysis excludes qualitative and non-financial factors, such as customer satisfaction or regulatory compliance, which could provide critical context. In addition, due to data constraints, reliance on common financial ratios excluded other essential measures, such as Return on Sales (ROS), Return on Assets (ROA), and Return on Equity (ROE). The study's focus on the medical insurance sector further limits the generalizability of its findings to the broader insurance industry.
From a methodological perspective, the use of LSTM networks for dimensionality reduction and temporal pattern extraction proved effective in capturing meaningful insights from time-series financial data. However, while LSTM excel at modeling sequential dependencies and preserving temporal relationships, their technique can limit interpretability compared to traditional methods such as Principal Component Analysis (PCA) or Gaussian Mixture Models (GMM). Additionally, LSTM require large datasets for optimal training, making them sensitive to data availability and quality. Future studies could explore hybrid approaches, such as integrating LSTM with attention mechanisms, to enhance interpretability and improve performance in cases with limited financial data.

The choice of scaling technique (either within-company or global scaling) remains a critical factor in determining the quality and interpretability of clustering results. Within-company scaling ensures that the variability within individual companies is emphasized, while global scaling standardizes data across all companies, highlighting relative differences. These strategies significantly impact the clustering process and may lead to varying interpretations of company trajectories and relationships. Further investigation into the implications of scaling decisions could yield insights into optimizing clustering frameworks for financial datasets.

Data gaps present another challenge, particularly in clusters characterized by inconsistent reporting. Variations in reporting frequencies or missing data can distort temporal analyses and complicate cross-cluster comparisons. Addressing these inconsistencies through imputation techniques or enhanced data collection practices could improve the reliability and robustness of clustering outcomes.
To build on the findings of this study, future research should expand the analytical scope to include non-financial metrics, such as customer satisfaction, regulatory compliance, and competitive positioning. Incorporating granular data, such as company-level strategic decisions or macroeconomic indicators, could provide deeper insights into performance drivers. Additionally, exploring advanced machine learning models like probabilistic clustering methods could reveal complementary perspectives and improve predictive accuracy. Addressing data inconsistencies in clusters with irregular reporting patterns, particularly Clusters 3 and 4, could result in actionable recommendations for enhancing governance and transparency in the sector.
By advancing the methodologies and addressing data challenges highlighted in this study, future research can contribute to a more comprehensive understanding of financial performance dynamics within the insurance industry. These efforts will support the development of robust, data-driven strategies for improving resilience and market transparency.

\section{Conclusion}
This study identified four distinct clusters of insurance companies based on their financial performance and reporting patterns, leveraging LSTM networks for dimensionality reduction and K-Means and Hierarchical clustering for pattern identification. The application of LSTM enabled the extraction of meaningful temporal trends, allowing for the grouping of companies with similar financial trajectories. 
Clusters 1 and 2 comprise companies with relatively stable performance, with Cluster 2 demonstrating resilience despite occasional fluctuations, while Cluster 1 reflects steady improvements in later stages. In contrast, Clusters 3 and 4 reveal irregular reporting behaviors, where data discontinuities and delayed reporting create challenges in assessing long-term financial stability. 
While these findings offer valuable insights into the dynamics of the insurance sector, they also highlight the critical need to address data gaps and refine scaling techniques to enhance clustering accuracy. Future research could integrate non-financial metrics and explore advanced analytical methods to deepen the understanding of financial performance and reporting behaviors. By addressing these limitations, future studies can provide a more comprehensive and nuanced perspective on the factors influencing stability and transparency in the insurance industry.

\section{Ethics approval and consent to participate}
Not applicable.

\section{Consent for publication}
Not applicable.

\section{Data Availability}
Code and data used for this analysis are available on GitHub in \cite{code}.

\section{Competing interests}
The authors declare that they have no competing interests.

\section{Funding}\label{sec8}
No financial support was received for this study.

\end{document}